\documentclass[twocolumn,epjc3]{svjour3}
\smartqed  

\usepackage{graphics}
\usepackage[space]{grffile}
\usepackage{latexsym}
\usepackage{url}
\usepackage[utf8]{inputenc}
\usepackage{fancyref}
\usepackage{hyperref}
\hypersetup{colorlinks=false,pdfborder={0 0 0},}
\usepackage{textcomp}
\usepackage{longtable}
\usepackage{multirow,booktabs}
\usepackage{lineno}
\usepackage{color}
\usepackage{mathptmx}      
\usepackage{amsmath}
\usepackage{enumitem}

\journalname{Eur. Phys. J. C}
\begin{document}
\title{The KASCADE Cosmic-ray Data Centre KCDC: Granting Open Access to Astroparticle Physics Research Data}

\author{
A~Haungs\thanksref{1,corr}  \and
D~Kang\thanksref{1}  \and
S~Schoo\thanksref{1,t1}   \and
D~Wochele\thanksref{1}  \and
J~Wochele\thanksref{1,corr1}  \and
W~D~Apel\thanksref{1}  \and
J~C~Arteaga-Vel\'azquez\thanksref{2}  \and
K~Bekk\thanksref{1}  \and
M~Bertaina\thanksref{3}  \and
J~Bl\"umer\thanksref{1,t2}  \and
H~Bozdog\thanksref{1}  \and
I~M~Brancus\thanksref{4,t3}   \and
E~Cantoni\thanksref{3,t4}  \and
A~Chiavassa\thanksref{3}  \and
F~Cossavella\thanksref{1,t5}  \and
K~Daumiller\thanksref{1}  \and
V~de Souza\thanksref{6}  \and
F~Di~Pierro\thanksref{3}  \and
P~Doll\thanksref{1}  \and
R~Engel\thanksref{1}  \and
B~Fuchs\thanksref{1,t6}  \and
D~Fuhrmann\thanksref{7,t7}  \and
A~Gherghel-Lascu\thanksref{4}  \and
H~J~Gils\thanksref{1}  \and
R~Glasstetter\thanksref{7}  \and
C~Grupen\thanksref{8}  \and
D~Heck\thanksref{1}  \and
J~R~H\"orandel\thanksref{9}  \and
T~Huege\thanksref{1}  \and
K~H~Kampert\thanksref{6}  \and
H~O~Klages\thanksref{1}  \and
K~Link\thanksref{1}  \and
P~{\L}uczak\thanksref{10}  \and
H~J~Mathes\thanksref{1}  \and
H~J~Mayer\thanksref{1}  \and
J~Milke\thanksref{1}  \and
B~Mitrica\thanksref{3}  \and
C~Morello\thanksref{5}  \and
J~Oehlschl\"ager\thanksref{1}  \and
S~Ostapchenko\thanksref{11}  \and
M~Petcu\thanksref{4}  \and
T~Pierog\thanksref{1}  \and
H~Rebel\thanksref{1}  \and
M~Roth\thanksref{1}  \and
H~Schieler\thanksref{1}  \and
F~G~Schr\"oder\thanksref{1}  \and
O~Sima\thanksref{12}  \and
G~Toma\thanksref{4}  \and
G~C~Trinchero\thanksref{5}  \and
H~Ulrich\thanksref{1}  \and
A~Weindl\thanksref{1}  \and
J~Zabierowski\thanksref{10}
}

\thankstext{corr}{e-mail: andreas.haungs@kit.edu}
\thankstext{t1}{Present address:  Siemens AG, Germany}
\thankstext{corr1}{e-mail: juergen.wochele@kit.edu}
\thankstext{t2}{now Head of KIT Division V - Physics and Mathematics}
\thankstext{t3}{deceased}
\thankstext{t4}{Present address:  INRIM, Torino, Italy}
\thankstext{t5}{Present address:  DLR Oberpfaffenhofen, Germany}
\thankstext{t6}{Present address:  DLR Stuttgart, Germany}
\thankstext{t7}{Present address:  University of Duisburg-Essen, Germany}

\institute{
Institut f\"ur Kernphysik \& Institut f\"ur Experimentelle Teilchenphysik, KIT - Karlsruher Institut f\"ur Technologie, Germany\label{1} \and
Universidad Michoacana, Instituto de F\'isica y Matem\'aticas, Mexico\label{2} \and
Dipartimento di Fisica, Universit\`a degli Studi di Torino, Italy\label{3} \and
National Institute of Physics and Nuclear Engineering, Bucharest, Romania\label{4} \and
Istituto di Fisica dello Spazio Interplanetario, INAF Torino, Italy\label{5} \and
Universidade S\~ao Paulo, Instituto de F\'{\i}sica de S\~ao Carlos, Brasil\label{6} \and
Fachbereich Physik, Universit\"at Wuppertal, Germany\label{7} \and
Fachbereich Physik, Universit\"at Siegen, Germany\label{8} \and
Dept. of Astrophysics, Radboud University Nijmegen, The Netherlands\label{9} \and
National Centre for Nuclear Research, Department of Astrophysics, {\L}\'{o}d\'{z}, Poland\label{10} \and
Frankfurt Institute for Advanced Studies (FIAS), Frankfurt am Main, Germany\label{11} \and
Department of Physics, University of Bucharest, Bucharest, Romania\label{12}
}

\date{Received: date / Revised version: date}

\maketitle

\begin{abstract}
The `KASCADE Cosmic ray Data Centre' is a web portal (\url{https://kcdc.ikp.kit.edu}),
where the data of the astroparticle
physics experiment KASCADE-Grande are made available for the interested public.
The KASCADE experiment was a large-area detector for the
measurement of high-energy cosmic rays via the detection of extensive air showers.
The multi-detector installations KASCADE and its extension KASCADE-Grande stopped the
active data acquisition
in 2013 
after more than 20 years of data taking.
In several updates since our first release in 2013 with KCDC we provide the public
measured and reconstructed parameters of more than 433 million air showers.
In addition, KCDC provides meta data information and documentation to enable a user
outside the community of experts to perform their own
data analysis. Simulation data from three different high energy interaction models
have been made available as well as a compilation of measured and published spectra
from various experiments. In addition, detailed educational examples shall encourage
high-school students and early stage researchers to learn about astroparticle physics,
cosmic radiation as well as the handling of Big Data and about the sustainable and
public provision of scientific data.
 \end{abstract}

\modulolinenumbers[5]

\section{Introduction}\label{introduction}

\begin{figure}[t]
\centering
\resizebox{0.3\textwidth}{!}{%
\includegraphics{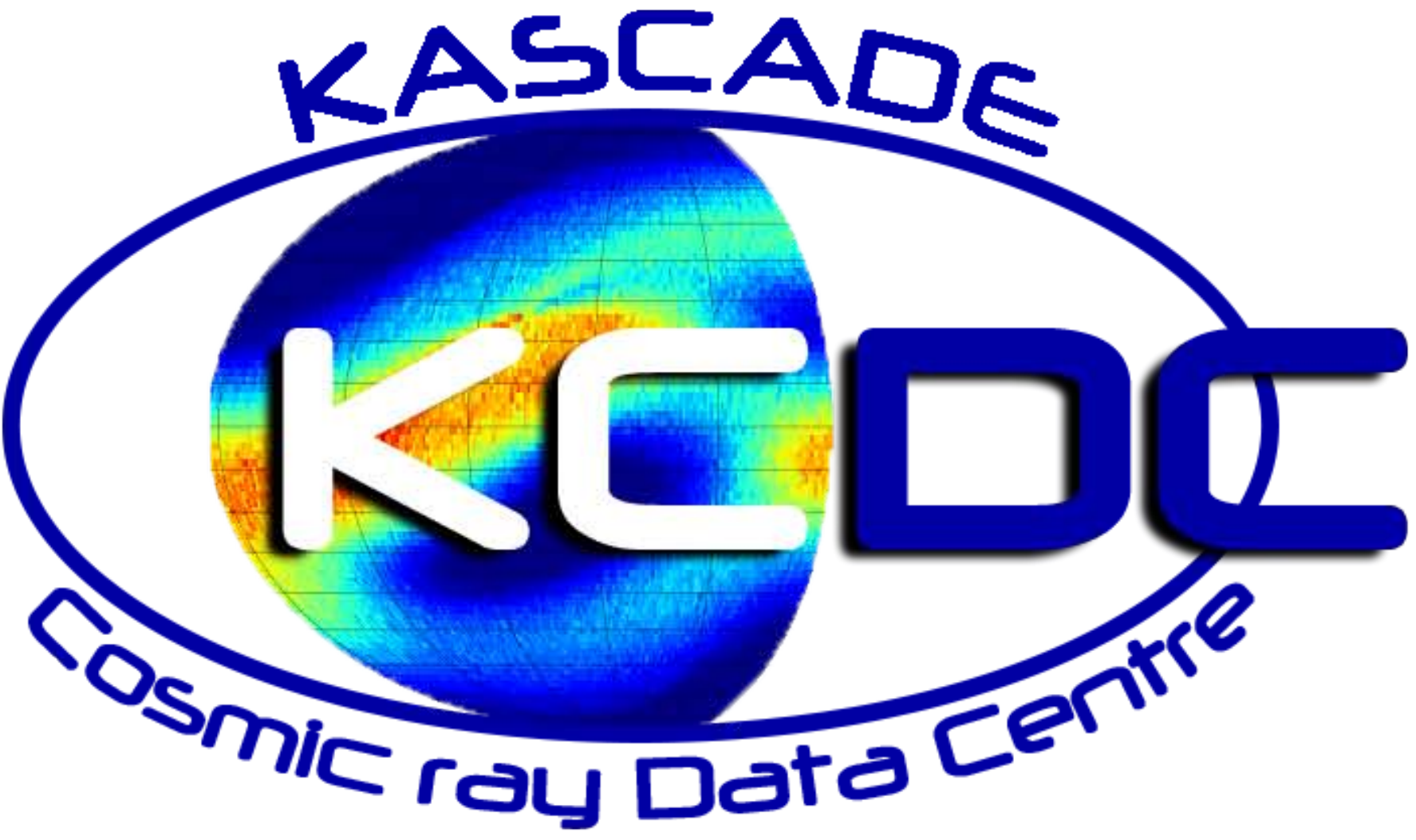}
}
\caption{Logo of KCDC (\url{https://kcdc.ikp.kit.edu}).}
\label{fig_logo}
\end{figure}

The \textbf{K}ASCADE \textbf{C}osmic ray \textbf{D}ata \textbf{C}entre
(KCDC) provides access to the collected cosmic-ray data of the KASCADE
and the KASCADE-Grande experiments.
In this paper we aim to introduce KCDC (logo see~fig.~\ref{fig_logo}), where
this publication will serve as the reference for users of the data provided by
KCDC.
We lay out
the motivation behind and the need for such a kind of public data
source and we discuss the available data sets.
Furthermore we present the advantages for physicists and non-scientists
alike using the data source and the advantages KCDC offers
for open data publications.
We describe shortly the evolution of KCDC and give
finally an outlook on possible use-case analyses
for the available data set and on the future of the KCDC project itself.

A first release~\cite{Haungs:2015zna} of KCDC is running since November 2013 with a
positive response from the community and public users.
Motivated by this success, we had several KCDC updates,
the last major release, called NABOO 1.0, in February 2017 and NABOO 2.0 and 2.1 in October
2017 and March 2018, respectively. Presently we provide in different formats data
from more than 433 million events from the three detector components KASCADE
(representing the original KASCADE Array), the Central Hadron Calorimeter,
and the array of the extension KASCADE-Grande. With the latest updates we provide as well
simulations, separately for the three detector components for direct download
as ROOT files and the data points of 88 published spectra from 21 experiments.

\section{KASCADE-Grande}

\begin{figure}[h]
\centering
\resizebox{0.44\textwidth}{!}{%
\includegraphics{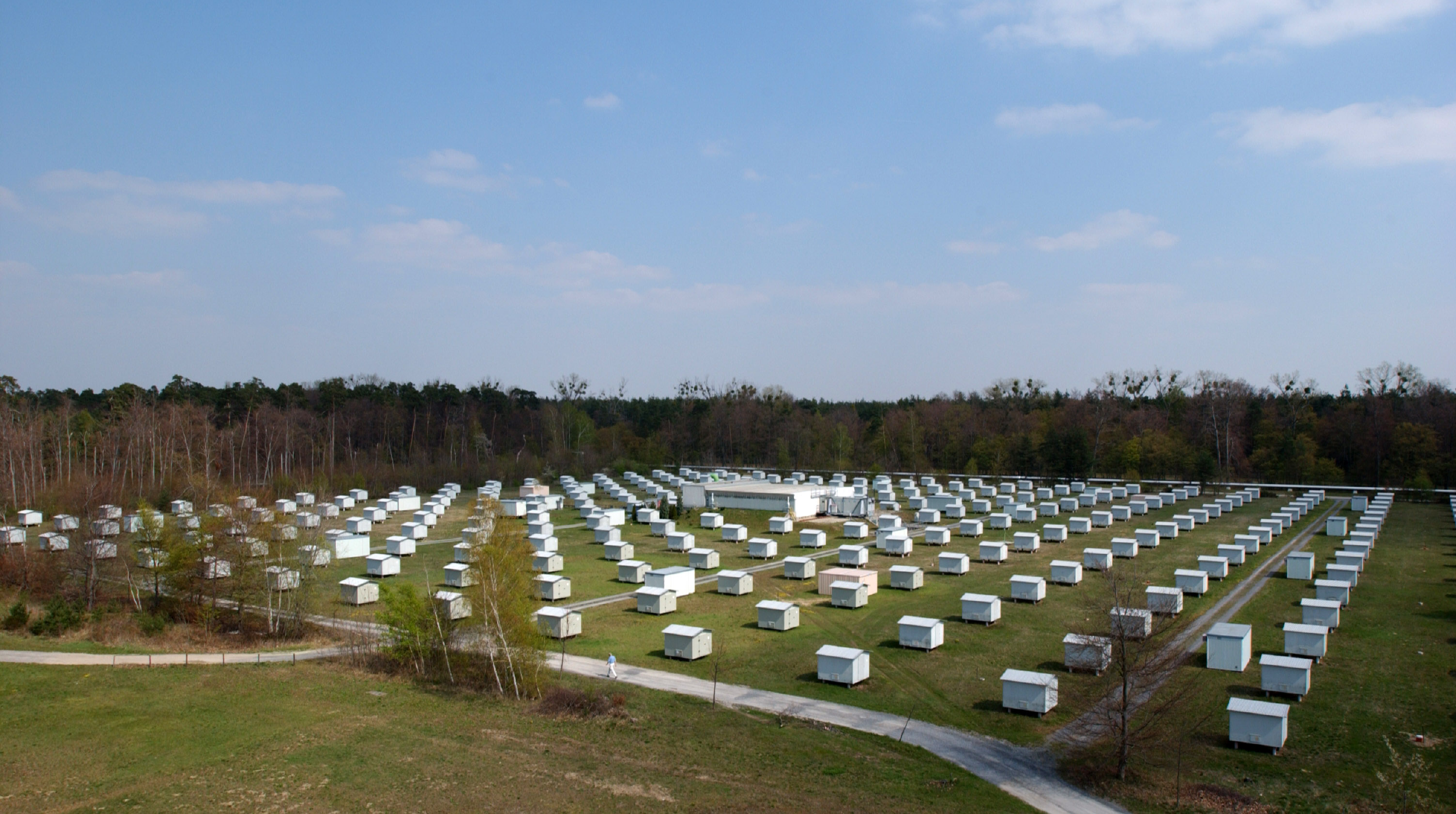}
}
\centering
\resizebox{0.45\textwidth}{!}{%
\includegraphics{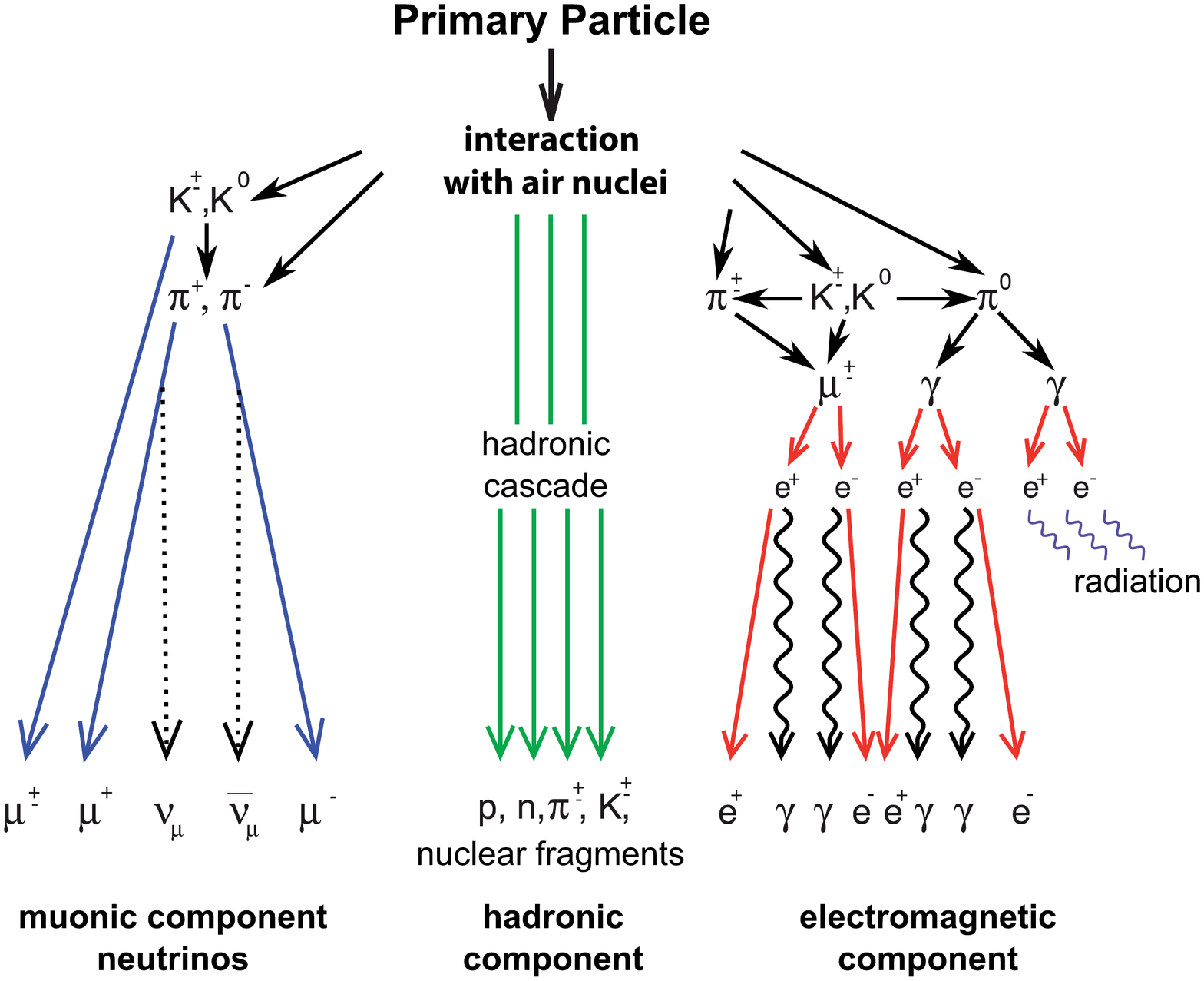}
}
\caption{Photograph of the KASCADE array with its central detector building (upper panel);
Schematic view of an extensive air shower (EAS), where KASCADE is measuring the hadron,
muon, and electron components (lower panel).}
\label{fig_kascade}
\end{figure}

KASCADE-Grande (KArlsruhe Shower Core and Array DEtector with its extension Grande)
was an extensive air shower experiment array to study the cosmic ray
primary composition and the hadronic interactions in the energy range
$E_0=10^{14}{-}{10}^{18} \: {\rm eV}$ (fig.~\ref{fig_kascade}).
The experiment was situated on site of the KIT,
Campus North (the former Forschungszentrum Karlsruhe) ($49.1 \: ^{\circ} {\rm N }$,
$8.4 \: ^{\circ} {\rm E}$) at $110 \: {\rm m}$ asl, corresponding to an average atmospheric
depth of $1022 \: \rm g/cm^{2}$~\cite{kascade}.
The main detector components of KASCADE-Grande were the KASCADE array (1996-2012),
the Muon Tracking Detector and the Central Detector to measure the hadronic and muonic
components in the center of the showers as well as
the extension Grande (2003-2012)~\cite{grande} to enlarge the detector area by a factor of
10 and to extend the accessible energy range to $E_0=10^{18} \: {\rm eV}$.

The radio antenna field LOPES~\cite{lopes} and the
microwave experiment CROME~\cite{crome} were also important
components of the experimental set-up of KASCADE-Grande, where the data
are not yet included in KCDC.
The full facility was in operation until end of 2012.
In this section we give a short introduction to the experimental set-up of KASCADE-Grande,
its main goals, and achievements of the 20-year running period.

\subsection{Experimental set-up}

\subsubsection{The KASCADE array}

\begin{figure}
\centering
\resizebox{0.45\textwidth}{!}{%
\includegraphics{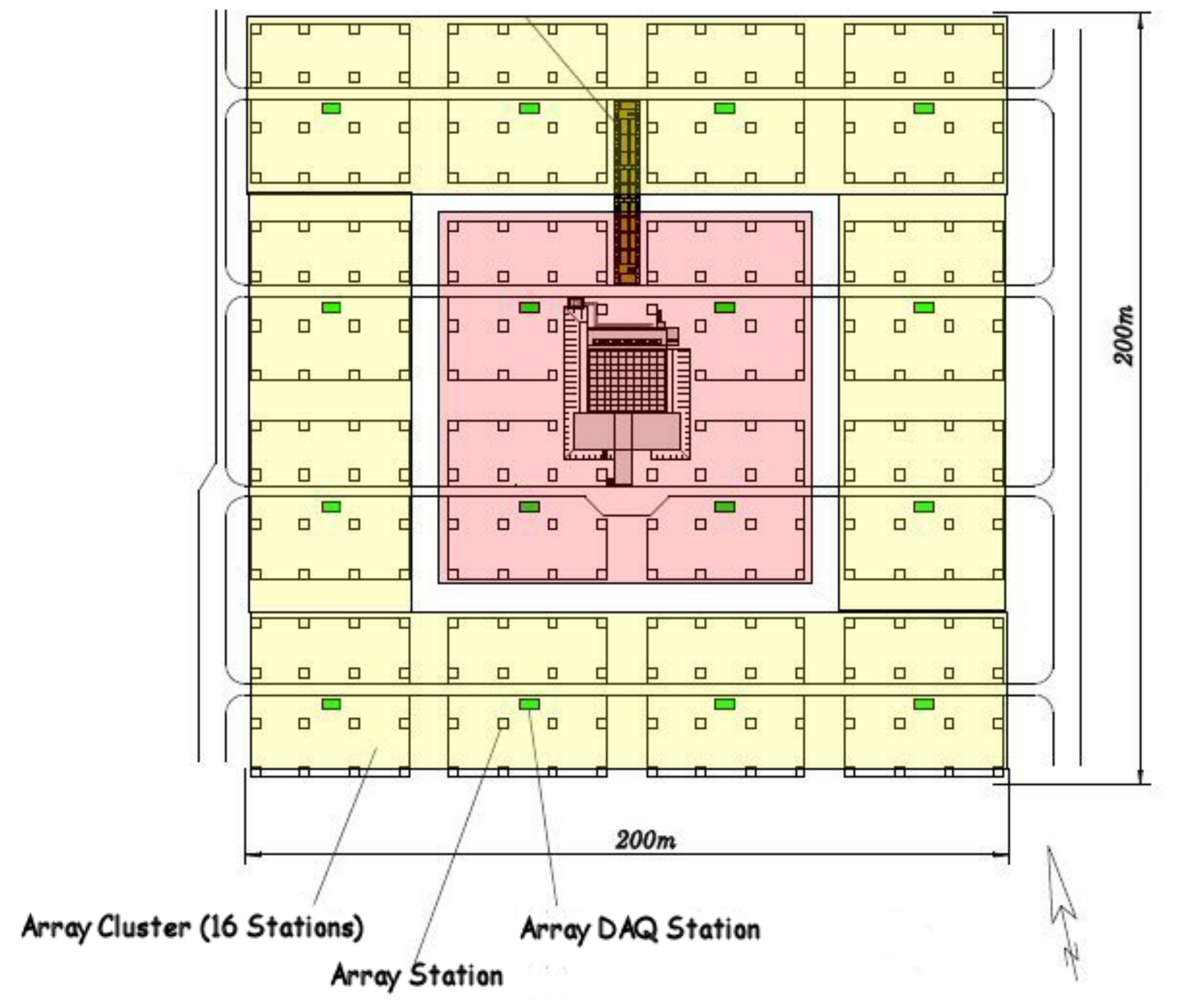}
}
\caption{Layout of the KASCADE array with the muon underground tunnel and the Central Detector of KASCADE.}
\label{ArrayLayout}
\end{figure}

\begin{figure}
\centering
\resizebox{0.45\textwidth}{!}{%
  \includegraphics{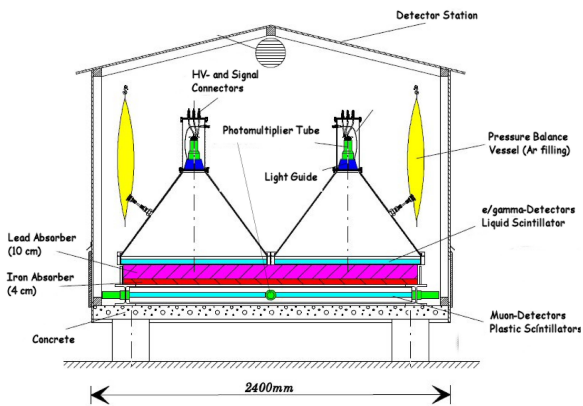}
}
\caption{The KASCADE array detector station.}
\label{ArrayStation}
\end{figure}

The KASCADE array consisted of $252$ scintillator detector stations set up in a regular grid with
$13 \: {\rm m}$ spacing covering an area of $200\times 200 \:{\rm m^2}$.
The stations are organized in $16$ clusters of $4 \times 4$ stations each (fig.~\ref{ArrayLayout}).
The stations of the inner 4 clusters contains 4 unshielded liquid
scintillation detectors ($e/\gamma$ detectors) each, to measure the charge particle
component and the particle arrival times.
The outer $12$ cluster consists of 2 liquid scintillation detectors only, but have in addition
lead and iron absorber sheets ($10 \: {\rm cm}$ Pb and $4 \: {\rm cm}$ Fe) underneath the
$e/\gamma$ detectors to measure the muonic shower component (fig.~\ref{ArrayStation}).
Here, vertical muons have a threshold of $230 \: {\rm MeV}$. Data are accumulated in the electronic
station of each cluster independently and transmitted to the data acquisition system (DAQ).

\subsubsection{The KASCADE Central Detector}

The central detector located in the centre of the KASCADE array covered an area of
$16 \times 20 \: {\rm m^2}$ and housed several detector components.
The main part was the finely segmented hadron sampling calorimeter~\cite{calorimeter} to detect the
hadronic component of an extensive air shower in about 11,000 warm-liquid ionization chambers
filled with Tetra\-methyl\-silane (TMS) or Tetra\-methyl\-pentane (TMP) and
mounted between layers of iron absorbers. The thickness of the calorimeter corresponds to
$11.5$ nuclear interaction lengths, so that hadrons up to $25 \: {\rm TeV}$ were absorbed
with less then $2.5{\rm \%}$ energy leakage.

In the third gap from the top of the iron stack a layer of $456$ scintillation detectors with a
size of $0.45 \: {\rm m^2}$ was installed. This layer was mainly used for fast triggering.
$50$ scintillation detectors of the same type were placed on the
top of the calorimeter.
Below the iron stack layers of position sensitive gaseous detectors were used for the
measurement of muon tracks.
Data of these devices as well as data from the underground Muon Tracking
Detector~\cite{myontrackdet} placed in
a tunnel north of the Central Detector are not yet included in KCDC.

\subsubsection{The Grande array}
\label{Grande1}

\begin{figure}
\centering
\resizebox{0.45\textwidth}{!}{%
\includegraphics{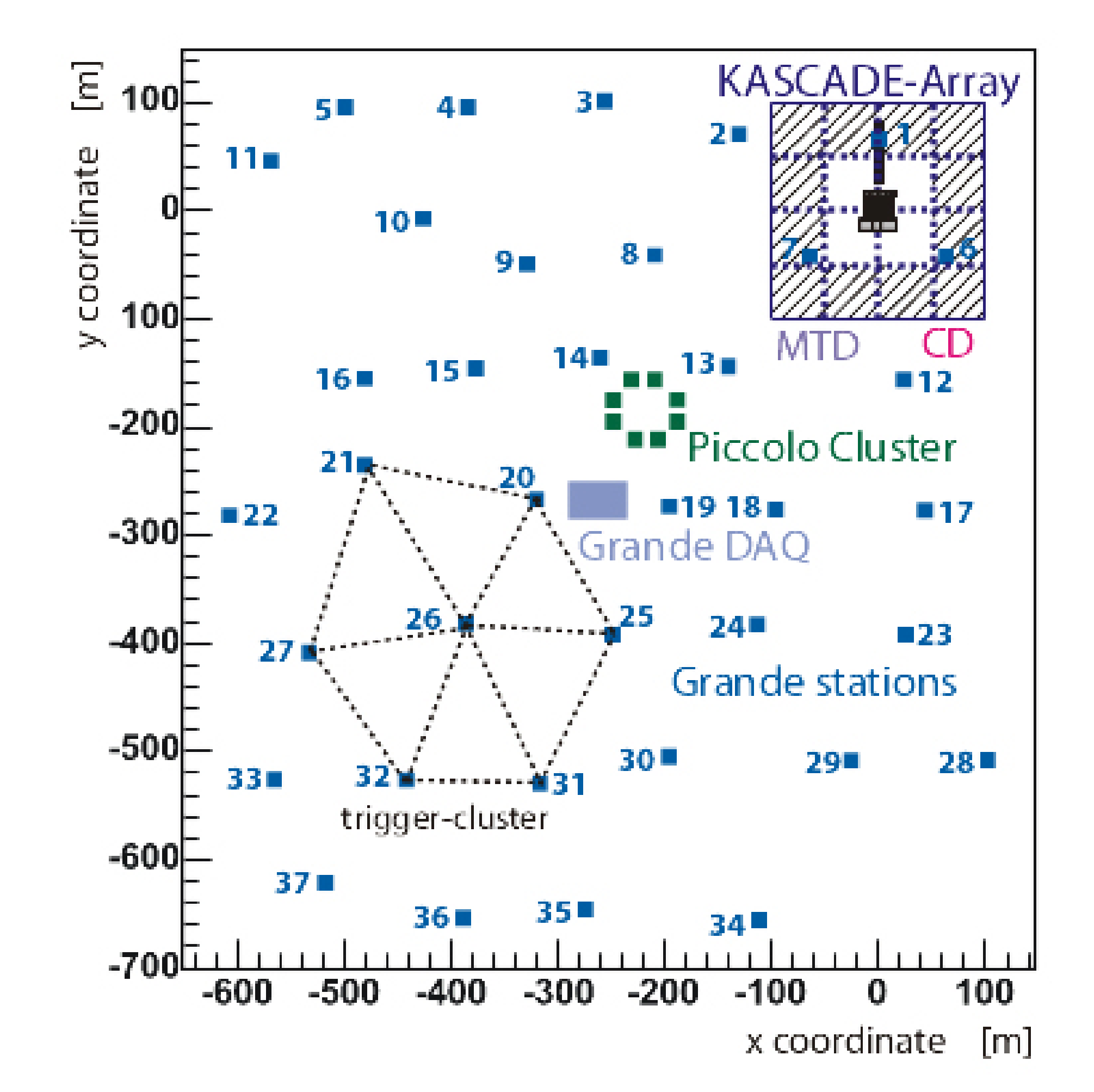}
}
\caption{The Grande Array layout.}
\label{Grande}
\end{figure}

During 2003 the Grande array was added to the KASCADE experiment extending its effective area,
and thereby its upper energy limit, by a factor of ten. The Grande array~\cite{grande}
consisted of 37 detector stations installed in an irregular triangular grid with an average
spacing of $137\,$m.
It covered an area of approximately $0.5\,$km$^{2}$ (fig.~\ref{Grande}).
Each detector station included a total of $10\,$m$^{2}$ of plastic scintillator subdivided
into 16 individual modules each viewed from the bottom by a photomultiplier.
The Grande array operated independently from the KASCADE array and measured the same parameters
of the charged particle component. The 37 Grande stations were organized in 18 overlapping trigger
clusters, where each cluster included seven detector stations: six in a hexagonal shape and a central one.
The data acquisition is triggered by either a 4/7 coincidence from a cluster (4 stations in a compact
configuration, trigger rate circa $5\,$Hz) or by a central trigger coming from KASCADE
(circa $3.5\,$Hz).
In addition, any full 7/7 coincidence (circa $0.5\,$Hz) from one of the 18 clusters is transmitted
also to KASCADE as a KASCADE-Grande trigger to start KASCADE-Grande event acquisition.
The Grande data analysis is in a first step performed by an independent processor in the
reconstruction program.

\subsection{The physics of KASCADE-Grande}

The main goal of the measurements was the estimation of energy and
mass of the primary particles in a wide energy range. The analysis is based
on the combined investigation of the charged particle, the electron,
and the muon components measured by the detector arrays of KASCADE
and Grande. The excellent timing of the detectors leads to an angular resolution
of about $0.1\,^{\circ}$ in shower direction,
which enables us to search also for large-scale anisotropies
as well as for cosmic ray point sources.

The general idea of the data analyses of the KASCADE and KASCADE-Grande
experiments is the determination of the chemical composition of cosmic rays in the primary
energy range $10^{14} - 10^{18}\,$eV by reconstructing individual
mass group spectra. Structures observed in these individual
spectra provide strong constraints to astrophysical models of
origin and propagation of high-energy cosmic rays to reach a better understanding
of energetic processes in our Universe.
For the interpretation of air-shower measurements in terms of energy and mass of the
particle or nucleus entering our atmosphere, models are in use, which
describe the interactions at energies similar and higher than reachable in man-made accelerators.
Systematic uncertainties due to these models are still the greatest obstacle in
understanding cosmic radiation.

\subsubsection{The all-particle energy spectrum}

\begin{figure}
\centering
\resizebox{0.49\textwidth}{!}{%
\includegraphics{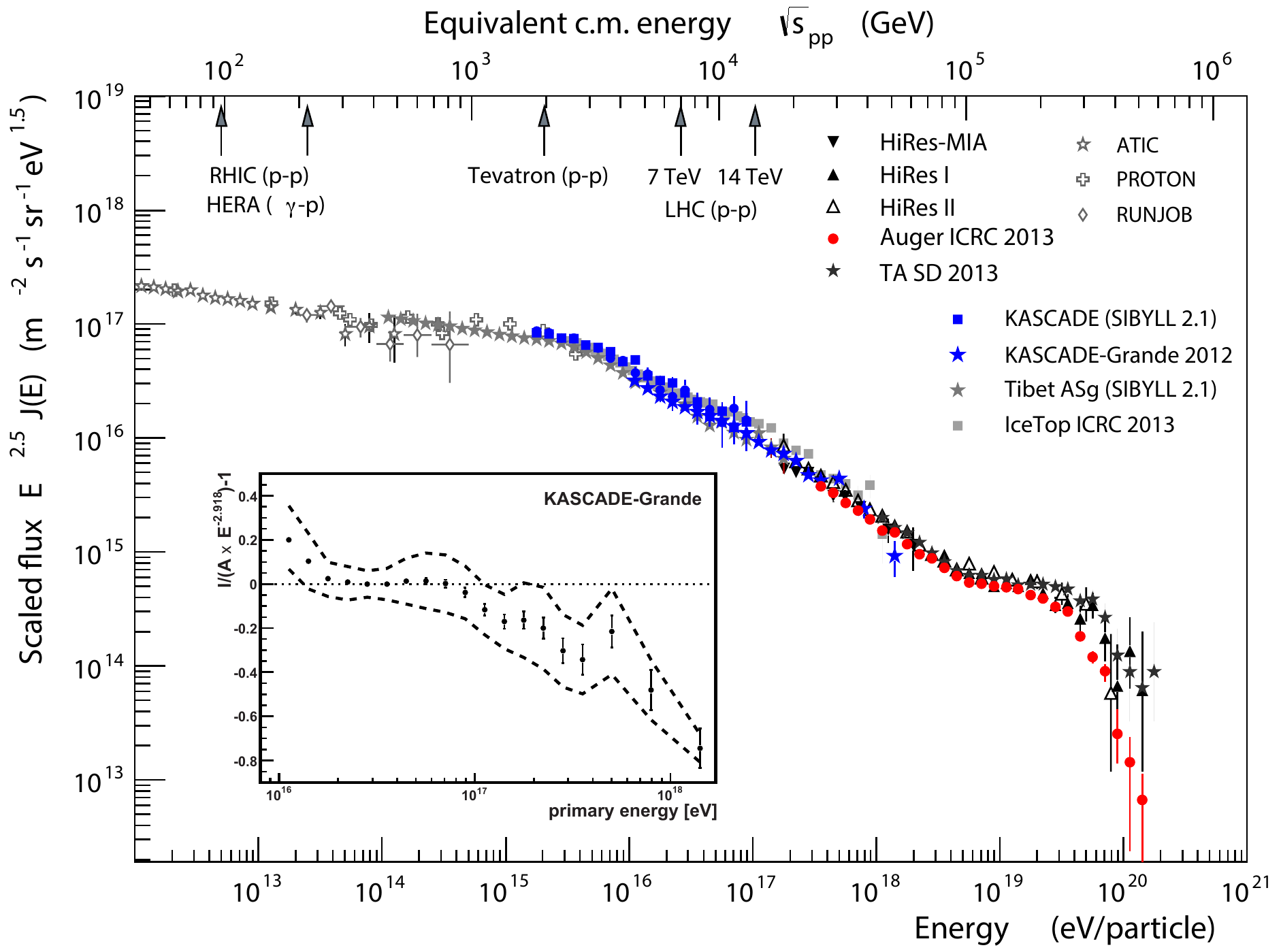}
}
\caption{The all-particle energy spectrum~\cite{KGhaungs-ICRC15} obtained with KASCADE
        and KASCADE-Grande (based on the QGSJet-II model and unfolded,
        i.e. corrected for the reconstruction uncertainties).
        Shown are the spectra in comparison with results of other
        experiments. In addition, the corresponding interaction energy
        at accelerators are indicated.
        The inlet shows the residuals of the all-particle energy spectrum from
        KASCADE-Grande with systematic uncertainties.}
\label{fig:spec}
\end{figure}

By using the specific hadronic interaction model QGSJet-II as a baseline,
a composition independent all-particle energy spectrum of cosmic rays was
reconstructed in the energy range from $10^{16}\,$eV to $10^{18}\,$eV
from the data of the Grande extension~\cite{Apel2012183}.
The spectrum is in the overlapping region in agreement
with the earlier published spectrum by KASCADE~\cite{kas-unf}
in the range of $10^{15}\,$eV to $10^{17}\,$eV.
Significant structures are observed in the all-particle spectrum
(fig.~\ref{fig:spec}):
The justification of the `knee' at a few times $10^{15}\,$eV
is given since many years (see ref.~\cite{haungs-review} and references therein).
In addition, with KASCADE-Grande, there is now a
clear evidence that just above $10^{16}\,$eV the spectrum shows a
significant `concave' behavior.
A further feature in the spectrum is a small break,
i.e.~knee-like feature at around $10^{17}\,$eV.
Found first by KASCADE-Grande
this is meanwhile confirmed by other experiments.
This `second knee' occurs at an energy where the rigidity
dependent, i.e. charge dependent, knee of the iron component
would be expected, if the `first knee' is caused by light primaries.
The concave part of the spectrum is then a consequence of knee-like features
of the spectra of medium masses.

\subsubsection{Elemental composition of cosmic rays}

\begin{figure}
\centering
\resizebox{0.45\textwidth}{!}{%
\includegraphics{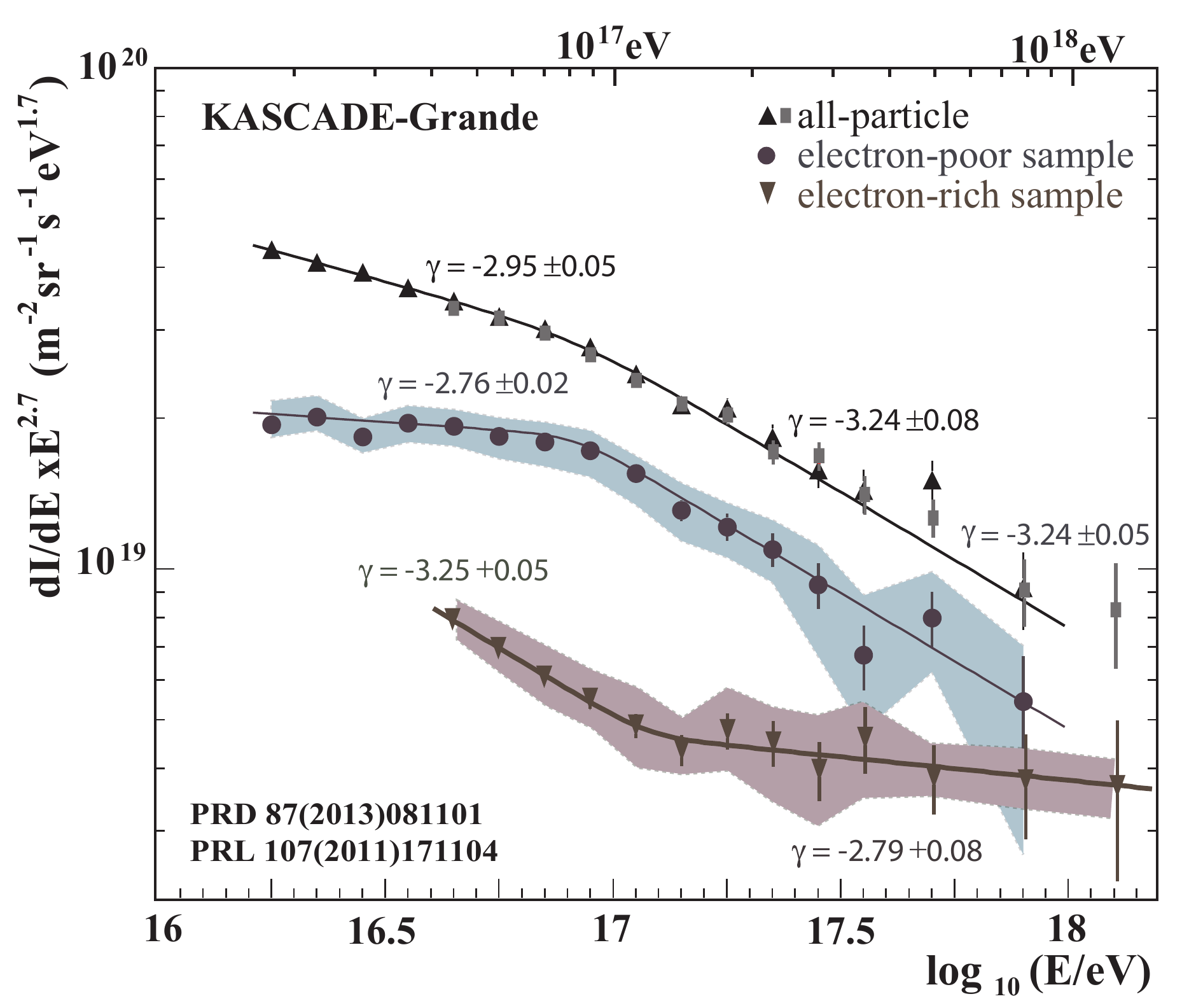}
}
\caption{All-particle, electron-poor, and electron-rich
energy spectra from KASCADE-Grande~\cite{KGhaungs-ICRC15}}.
\label{fig:comp}
\end{figure}

Already in 2005 KASCADE demonstrated~\cite{kas-unf}
that the knee is caused by a decrease of the light mass group
of primary particles and not by medium and heavy primary particles.
With KASCADE-Grande we investigated such individual mass group
spectra also at higher primary energies~\cite{prl107,Apel2013}.
The application of this methodical approach to shower selection and
separation in various mass groups were performed and cross-checked
in different ways, where figure~\ref{fig:comp} shows the
main results. In a first step we separated the Grande data in two mass
groups only, i.e., in groups of heavy and light primary masses according to the
electron-muon ratio in the air showers.

The reconstructed spectrum of the electron-poor events,
i.e. the spectrum of heavy primaries, shows a distinct knee-like
feature at about $8 \cdot 10^{16}\,$eV~\cite{prl107}.
The selection of heavy primaries enhances the knee-like feature that is
already present in the all-particle spectrum.
Despite the fact, that the relative abundance of the heavy particles
varies significantly depending on the model in use, the spectral feature
of this `heavy' knee is visible in all the electron-poor spectra.
In addition, an ankle-like feature was observed in the spectrum
of the electron-rich events, i.e.~light elements of the primary cosmic rays,
at an energy
of  $10^{17.08 \pm 0.08} \, \mathrm{eV}$, hinting to an occurrence of a
component of cosmic rays which have their origin in the extra-galactic space.

\subsubsection{Hadronic interaction models}

Historically, a great achievement of KASCADE was the invention of
the air-shower simulation tool CORSIKA
(\textbf{CO}smic \textbf{R}ay \textbf{SI}mulation for \textbf{KA}scade), which
meanwhile is used by all major air-shower experiments worldwide~\cite{corsika}.
Mainly data of the hadron calorimeter and additional muon counters at the
central detector and the muon tunnel were used within KASCADE to iteratively
test and improve the various versions of the hadronic interaction models
optionally available in CORSIKA (see e.g.,~\cite{hadrint,mtdint} and
references therein).
Regarding the analysis of KASCADE and KASCADE-Grande data one has to conclude
that all versions of hadronic interaction models of newest generation provide a
`physical' result in terms of energy and composition of primary cosmic rays.
Physical means here that the mean composition lies within the band spanned
by primary proton and iron simulations. However, the absolute energy and, in particular,
the mass scale varies significantly from model to model. In addition, it varies partly also
within the models if different shower observables taken from the electromagnetic,
muonic, or hadronic components are used.
This ambiguity can not be resolved by looking at one single
observable or experiment, only. This is one reason to preserve and provide the data
of KASCADE-Grande also for future analyses, now via KCDC.

\section{KCDC}\label{kcdc}

With KCDC, the \textbf{K}ASCADE \textbf{C}osmic Ray \textbf{D}ata
\textbf{C}entre~\cite{kcdc}, we successfully provide public access to
experimental cosmic-ray data.
Via a web portal physicists as well as non-scientists have easy and
convenient access to the high-quality cosmic-ray data collected by the
KASCADE and KASCADE-Grande experiments.
With our last release, named NABOO~\cite{kcdc-icrc17},
we provide more than 433 million events from the whole
measuring time of KASCADE-Grande. A high quality of the
data provided by KCDC is achieved by regular internal quality
tests and 20 years of collected knowledge and experience gathered
by the KASCADE-Grande collaboration.

Open access as described e.g.~by the Berlin Declaration~\cite{Berlin}
includes free, unlimited access to scientific data collected with
financial aid from public institutions. One underlying notion behind the
term `Open Access' is that for research paid by public funding the tax payer
has the right to have free access to the data. This also implicitly
includes a permanent nature of this access such that the data source and
access conditions do not vary or change over time. Therefore, once
published data can not be revoked and have to remain accessible.
KCDC follows this notation as well as wants to contribute to the
development of general principles in the reuse of scholarly data.
We follow the guidelines of the FAIR Data Principles~\cite{refFAIR},
where FAIR stands for: Findable, Accessible, Interoperable, Reusable.

Alongside free access, Open Data also demands the publication of meta
information and documentation. This documentation has to provide
interested third parties all information to understand, work with and
process the data. In the case of physics data, this
includes a thorough and transparent description of the detector, the
detection process and all physics background the analyses are based
on. For an experiment like KASCADE-Grande with a life time of 20 years
or more this is a task of monumental proportions which obviously has no
simple or quick solution.

With KCDC we made the first step by publishing the reconstructed physics
data of the experiment and of the calibrated entry at each individual
detector per event.
This is not the collected raw data acquired by the detector, but a way
step before the final reconstructed results.
The high level data takes
into account the collected knowledge of the collaboration regarding the
detector system and the involved physics necessary for reconstructing
the quantities we are most interested in, like the cosmic ray energy.
This levitates from the need to gather an in-depth knowledge
on these aspects before the user can reasonably use the data.
This also implies that the background information on the detector and
the KASCADE experiment is available. KCDC makes all the information
easily accessible and offers detailed descriptions of the available data
sets, including relevant physics background information necessary
for data analysis.
Furthermore, KCDC maintains analysis examples which include source code
examples in C++ and Python. This provides an impression on how to work
with the data to an interested non-scientific audience. The examples are
also valuable to physicists as they are an easy starting point and
template for analyzing the data. \\

The current web portal provides two ways to access the physics data from
KASCADE-Grande. On the one hand we provide predefined selections
(parts of the complete data set) which correspond to examples on the
web portal or is referred to by the KCDC web page.
On the other hand the user can also make his own selections from the
complete data set. For this purpose a web form provides methods to
select for example a specific energy range, only events arriving from
certain directions or time windows. These selection cuts are then
transmitted and processed in the backend on our servers. Depending on
the amount of data and complexity of the applied cuts the user can
download the compressed data selection via FTP download after
some time. For convenience users may also receive a notification email
as soon as selection jobs are processed and their data is ready.

The feature of ordering specific user defined selections of the data set
provides several benefits especially for physicists. First of all, the
common data source provided by KCDC and the unified selection interface
make it easy for scientists to share their data bases and compare their
analysis. The KCDC quality criteria and regular data integrity checks
offer also a high quality and reliable data source, which can be
consulted from all over the world at any time. The possibility of high
level cuts on the data set furthermore reduces the amount of data to be
transferred and maintained by the individual scientist and provides a
shared facility for doing these standard selections thus minimizing the
chance of human error. Finally, the long term availability guaranteed by
KCDC reduces the difficulty to repeat an analysis on the same data set
in the future as the specific data set has not to be stored locally but
can be retrieved any time as required.

\subsection{First releases and experiences}

The aim of KCDC from beginning was the installation and establishment of a public
data centre for high-energy astroparticle physics.
In the research field of astroparticle physics, such a data release is a novelty,
whereas the data publication in astronomy has been established for a long time.
Therefore, there were no completed concepts, how the data can be treated and processed so that
they are reasonably usable outside the collaboration.
The first goal of KCDC was therefore to make the data from the KASCADE experiment available to the
community and observe the acceptance by the community and the public.
\begin{figure}
\centering
\resizebox{0.45\textwidth}{!}{%
\includegraphics{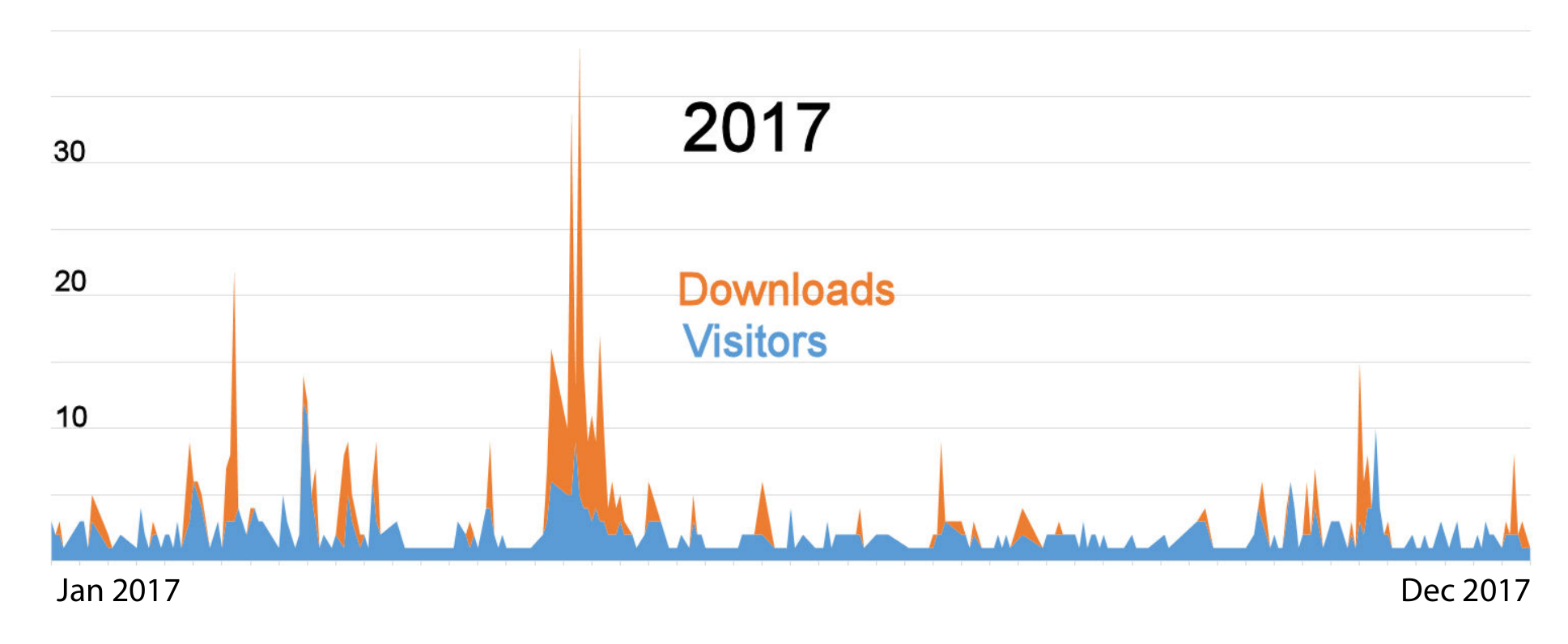}
}
\caption{KCDC access statistics for 2017. The peaks correspond to KCDC presentations at conferences.}
\label{monitoring}
\end{figure}
Already with the first release, KCDC provided efforts to fulfill following three basic requirements:
\begin{itemize}
\item {\bf KCDC as data provider:}
There is free and unlimited open access to KASCADE cosmic ray data, where a selection of fully
calibrated and reconstructed quantities per individual air shower is provided.
The access has to rely on a reliable data source with a guaranteed data quality.
\item {\bf KCDC as information platform:}
For a meaningful usage of KCDC, a detailed experiment description as well as sufficient
meta information on the provided data is needed for any kind of data analysis.
This is accompanied by a reasonable description of the physics background as well
as tutorials, focused for teachers and pupils.
\item {\bf KCDC as long-term digital data archive:}
To constitute a sustainable piece of work, KCDC serves also as archive of software and data
for the collaboration as well as for the public.
\end{itemize}
This concept of the data centre (software and hardware) is meanwhile implemented,
and was released as a series of KCDC versions.
However, the project faces still open questions, e.g. how to ensure a consistent calibration,
how to deal with data filtering and how to provide the data in a portable format as well as how a
sustainable storage solution can be implemented.
In addition, access rights and license policy play a non-negligible role and were considered in
detail, but needs to be reconsidered from version to version.

Since the first release (Open Beta Version) Wolf$\,359$, KCDC was further developed and
improved. Subsequently, the versions VULCAN, MERIDIAN and NABOO were released, where as
communication platforms serve an email list of the KCDC subscribers and social
networks, like Twitter (\url{https://twitter.com/KCDC_KIT}).
From beginning, a large interest from the community was given, proved, e.g. by our anonymous
monitoring of the access to the portal (see figure \ref{monitoring}) and the interest and
attendances at presentations at conferences or on public events (open days, science fairs, etc.).
In the first years of operation nearly 120 users registered from more than
30 countries distributed over 5 continents.

\subsection{Technical realization}\label{technical-realisation}

Adhering to the idea of open access for data, KCDC itself also only
relies on non-commercial and state of the art open source software. For
managing the webpages, data streams, data bases and communicating with
the backend, we focus on the web framework Django and Python 3
with several other open source libraries providing the backbone of the
main web service. Web pages are rendered in HTML by Django's template
engine enriched with our very own JavaScript (e.g.~the jQuery library)
and CSS additions. Interfaces to our data sources like a MongoDB database
for physics data are realized within Python.

The experimental data are stored in a noSQL database which enables us
to expand easily the number of events or of detector components
without the restraint of a fix database scheme.
We are using MongoDB 3.2 on a single server and are aiming at a
sharded cluster for better performance.
The full KCDC system runs on an Nginx server and communicates with the
database server and a worker node. On each
worker node, managed and monitored through Django via the Celery
extension, Python tools process the user selections.
The data packages reside on an FTP server, where they can
be retrieved by the users via an HTML link, provided when processing
of their job has been successfully finished.
Preselections can be directly accessed by a registered user via FTP.
To save storage space, each data package is only provided for two weeks.
After this time the data packages are automatically removed.
For easy access and information users can look up previously processed
selections in a user history page. There, the status of data processing for
requests, the exact details of the applied selection cuts and
additional information are provided.
This history can be managed by the user in the sense that requests can be
deleted or resubmitted with changes applied.

The web portal software and all accompanying tools are especially
designed with generality in mind. This includes a vast set of tools to
configure and manage the web portal directly via a web browser interface
without directly connecting to the server.
The KCDC software is also structured in a basic software package called
KAOS (Karlsruhe Astroparticle Open-data Software).
An installer is provided to setup the specific modules for an open access
web side which enables also other experiments to adapt the software to their own
needs. A plug-in system makes it easy to add functionality to the basic KAOS package
overwriting the KAOS default settings.

\section{Description of Datasets}
\label{description-of-datasets}

\subsection{Air-shower data}

The air showers measured by KASCADE-Grande are analyzed using the reconstruction program
KRETA (\textbf{K}ASCADE \textbf{R}econstruction for \textbf{E}x\textbf{T}ensive
\textbf{A}ir showers).
Starting from the energy deposits and the individual time stamps in all detectors
of all components KRETA determines physical quantities like the total number of
electrons, muons, hadrons in the shower or the shower direction.
KRETA reads the raw data, performs calibrations and reconstructs the
basic shower observables, and stores all results in histograms and vectors of
parameters.
KRETA is written in FORTRAN using CERN library packages and CERN’s HEPDB database
to hold time dependent calibration and status parameters for all detector components.

From the various observables obtained in the analysis we choose 22
for 158 million events to be published in the first revision of KCDC in
November 2013. With the version released early 2017 we published more
than 433 million events from the detector components KASCADE, Grande and Hadron
Calorimeter with 24 quantities on which cuts can be applied via the KCDC Data Shop and
5 arrays of data from the local station measurements.
Based on data generation and data handling we distinguish between:

\noindent \textbf{Measured Data} corresponding to data which
are directly measured or reconstructed by the KASCADE analysis software
like energy deposit and arrival directions.

\noindent \textbf{Calibration Data} used to calibrate and reconstruct the
data sets on an event-by-event basis like temperature and air pressure.

\noindent \textbf{Event Information} used to uniquely characterize an event like
event time and run or event number. \\

What follows is a brief description of the available KCDC quantities and procedures
of how they were obtained:  \\
\textbf{Primary energy (KASCADE):}
One of the main goals of the air shower measurement is to determine the energy
spectrum of the cosmic rays. Due to uncertainties in the hadronic interactions
and the shower-to-shower fluctuations due to the stochastic process in the shower development,
the determination of energy and mass is challenging.
In KASCADE, we measure the electromagnetic and muonic
components of air showers separately. By using both observables, we perform a transformation
matrix in order to convert the number of electrons and muons to the energy of primary particles
taking into account the angle-of-incidence. The parameters of the formula of the energy estimator
are derived from air-shower simulations using the simulation program CORSIKA
applying the hadronic interaction model QGSjet-II-02
(Quark-Gluon-String Model, version II-02~\cite{qgsjet})
for laboratory energies above $200\,$GeV and the low energy
model FLUKA 2002.4~\cite{fluka} below.
This first order rough energy estimation is given by the formula:
\begin{equation*}
    \label{eq:E0}
    \begin{split}
        lg(E_0/\mathrm{GeV}) = & 1.93499 + 0.25788 \cdot lg(N_{e}) + 0.66704 \cdot lg(N_{\mu}) \\
        &+ 0.07507 \cdot lg(N_{e})^{2} + 0.09277 \cdot lg(N_{\mu})^{2} \\
        &- 0.16131 \cdot lg(N_{e}) \cdot lg(N_{\mu})
    \end{split}
\end{equation*}
where lg($N_{e}$) and lg($N_{\mu}$) are corrected for atmospheric depth and
angle-of-incidence. \\
\textbf{Shower core (KASCADE):}
The core position is the reconstructed location of the shower centre
derived from the energy deposits of each detector station of one event.
By means of a neural network algorithm
which combines high efficiency for the identification of the shower core with good
rejection capability for showers that fall outside the fiducial volume, the core
can be determined to a precision of about $1\,$m.
Extensive air showers with a core position outside the detector area have a great
probability for being incorrectly reconstructed. Therefore, it is recommended to cut
showers with a core distance larger than $91\rm m$ from the centre of the detector area. \\
\textbf{Shower direction (KASCADE):}
The KASCADE detectors measure the arrival time and the energy deposit of air shower
particles. The shower directions are determined by evaluating the arrival
times of the first particle in each detector station. To increase the accuracy,
the energy deposits are taken into account when the direction of the shower disk is
calculated in a second order correction. By this, an angular resolution of
$0.1^\circ$-$1^\circ$ depending on the shower size $N_{e}$ is reached.
The angular resolution drops significantly above
$\theta > 40^\circ$, caused by the fact that the reconstruction algorithms has
been fine-tuned to zenith angles below $40^\circ$.
In KASCADE coordinates, the zenith angle is measured against the vertical direction,
which means that $\theta = 0^\circ$ is pointing upwards and $90^\circ$ denotes a
horizontally arriving shower.
The azimuth is defined as an angle measured clockwise starting in northern direction
($90^\circ$ is east).
The regular local orientation of the KASCADE array at KIT had an offset of
about $+15^\circ$ against the real North, which is corrected for in the data analysis. \\
\textbf{Number of electrons and muons (KASCADE):}
In the 252 e/$\gamma$- and 192 $\mu$-detectors, electrons and muons as well as other
particles are registered.
In three steps the number of electrons and muons are reconstructed where the results
of the current iteration level serves as starting parameters for the next step.
Every level starts with a consistency check and the preparation of the data.
Signals inconsistent with those of the neighbouring detectors are discarded
as well as signals with a time stamp more than $200\,$ns from the shower front.
Then the measured energies are corrected for the inclination of the shower
axis and the lateral energy correction function is applied. Thereby the e/$\gamma$-
detector signals are corrected for the contributions from $\gamma$-particles and
the $\mu$-detector signals are corrected for electromagnetic
and hadronic punch through. Finally the corrected signals are converted
to particle numbers. Finally, the lateral distribution of the densities are  fitted with a
modified NKG-function and integrated to obtain the total particle numbers. \\
\textbf{Shower age (KASCADE):}
Contrary to variables like number of electrons or muons the value of the
age parameter has no absolute meaning, as it depends on the choice of the lateral
distribution function which is fitted to the shower data. It may also be called
lateral shape parameter because it describes the steepness of the lateral
electron density distribution. KASCADE uses a modified NKG-function to fit
the lateral shower shape.
A heavy primary particle with the same energy as a
light one gives rise to a flatter lateral distribution, as the shower starts earlier
in the atmosphere.  When reaching ground, the shower is "older", which gives
the age parameter its name. The age parameter therefore may help
(in combination with the ratio of number of electrons to muons) to
distinguish between primary particles of different mass. \\
\textbf{Energy deposits (KASCADE):}
The energy deposit in every KASCADE station is recorded separately for the
signals of the e/$\gamma$-detectors and $\mu$-detectors.
The energy deposits of the e/$\gamma$-detectors are used to calculate the shower core
position and the shower energy by means of a lateral density function fit.
The mean energy deposit of a minimum ionising particle
(mip) is about $12\,$MeV.
Energy deposits  equivalent up to 1250 mips can be detected linearly
with a threshold of roughly 1/4 mip ($3\,$MeV).
The mean energy deposit of a minimum ionizing particle in the $\mu$-detectors
is about $8\,$MeV, where energy deposits equivalent to 60 mips can be detected
linearly with a threshold of roughly 1/4 mip ($2\,$MeV).
The energy deposits of both detector types are derived from the
stored ADC values for each detector station by means of a calibration procedure where
the influences of electronics and cabling are included.
In KCDC, no cuts can be applied to these quantities. \\
\textbf{Arrival times (KASCADE):}
The first particle passing the threshold in every station produces a time stamp called
Arrival Time which is recorded separately from the 252 $e/\gamma$-detectors.
The measured arrival time in each station is calibrated for delays and response
times of the individual station and stored with a resolution of $1\,$ns.
Arrival times are mainly used to calculate the shower direction. \\
\textbf{Shower core (Grande):}
The Grande shower core position is reconstructed independently from
KASCADE using the energy deposits from the 37 Grande detector stations. The
reconstruction method is the same as for KASCADE. \\
\textbf{Shower direction (Grande):}
The shower direction in Grande is reconstructed basically in the same way as for KASCADE,
using the arrival times of the first particle from every Grande station, corrected with
the energy deposits of the charged particles. The angular resolution is
$0.8^\circ$ with a small dependence on the shower size $N_{ch}$. The reconstruction
algorithm has been fine-tuned to zenith angles below $40^\circ$. \\
\textbf{Number of charged particles and muons (Grande):}
From the measurements of the energy deposits in the Grande array stations the
total number of charged particles in the shower, i.e.~the shower size is reconstructed.
The reconstruction is performed similar to KASCADE but with different parameters.
The number of muons ($N_{\mu}$) is derived from the KASCADE detector stations
participating in the respective event with a simplified method.
As there are normally only few KASCADE detector stations with muon information
when Grande has been triggered, the number of detected muons is compared to a shower
with a normalised shower size.
The average value of the ratio of ‘measured muon number’/’expected muon number’
over all detectors are formed and stored as $N_{\mu}$. \\
\textbf{Shower age (Grande):}
Like KASCADE, Grande uses a modified NKG-function to fit the lateral shower shape. \\
\textbf{Energy deposits (Grande):}
The energy deposits of the charged particles are used to calculate the
shower core position and the shower energy using a lateral density function
fit similar to KASCADE but with different parameters. The energy deposits are
derived from the stored ADC values for each detector station by means of a
calibration procedure where the influences of electronics and cabling are included. \\
\textbf{Arrival times (Grande):}
The arrival time is the first time stamp at each detector
station that has been hit by a charged particle. \\
\textbf{Number of hadrons (Calorimeter):}
The hadrons and their interactions are important for the understanding of the
shower development within the atmosphere.
Due to the fine lateral segmentation and the read-out of the KASCADFE hadron
calorimeter, hadrons with an energy $E_{had} >20\,$GeV
can be measured. They can be separated from each other when the
distance of their axis is above $40\,$cm. The spatial resolution of the
calorimeter is about $11\,$cm and the energy resolution is 30\% for hadrons with $100\,$GeV
decreasing to 15\% for $E_{had}=25\,$TeV. \\
\textbf{Hadron energy sum (Calorimeter):}
The energy sum of all reconstructed hadrons ranges between $20\,$GeV corresponding to
the lower threshold, and about $10^{7}\,$GeV. \\
\textbf{Air temperature and air pressure:}
The condition of the Earth’s atmosphere has an influence on the development of
the extensive air showers and thus cannot be neglected, in particular for
anisotropy studies of cosmic rays.
The variation of the air pressure of about 1 hPa corresponds to a
change in the measured rate of about 1\%,  while the effect of the temperature is
significantly smaller. The fluctuation of the rate because of the pressure
variation can be up to 20\%.
The meteorological data are provided by the Institute of Meteorology and
Climate Research at KIT. The measurements of the temperature and the air pressure
are taken from sensors placed $2\,$m above ground level for the temperature readings
and $1.5\,$m above ground for the air pressure measurements on site of KIT, in
about $1\,$km distance from the KASCADE experiment.
All climate observables were recorded every 10 minutes. \\
\textbf{Event time:}
An event is stored when a pre-defined trigger condition of any detector component
is fulfilled. The time of the first trigger is stored as the `Event Time'
and distributed to all other detector components.
The event time (DateTime) is always given in UTC. As a redundant time information we use
in KASCADE the Unix Time, a system time stamp counting the number of seconds elapsed
since January, 1$^{st}$ 1970 (midnight UT), which is internally referenced as Global
Time (GT). To get a high precision time stamp the Micro Time information (MT) is used.
Based on the cycle of a $5\,$MHz clock which is reset and synchronized every second, we
obtain an accuracy of $\pm 200\,$ns for the event time. \\
\textbf{Run and event number:}
Run number and event number are two parameters which characterize an event uniquely.
They are always supplied with the data sets. A run is defined as a set of
events recorded under the same hardware conditions.
The event number starts at one for each run and is increased with every valid hardware
trigger which invokes data recording. Run numbers and event numbers are not necessarily
in increasing order for the selected event sample in KCDC.

\subsection{Simulations}

Analysing experimental data of air showers in terms of
parameters of the impinging primary particle or nucleus
requires a detailed theoretical modeling of the entire cascade.
This can only be achieved by Monte-Carlo calculations taking into
account all knowledge of particle interactions and decays.
With KASCADE we have not only reconstructed energy spectra
for five mass groups using 6 different high-energy hadronic interaction
models, but also tested the validity of these models by studying
correlations of various individual observables.
This helped the model builders to improve their models.
All the models are implemented in the CORSIKA simulation package.
CORSIKA has been written especially for KASCADE and extended
since then to become the standard simulation package in the field
of cosmic ray air shower simulations~\cite{corsika}.

\begin{figure}[b]
\centering
\resizebox{0.45\textwidth}{!}{%
\includegraphics{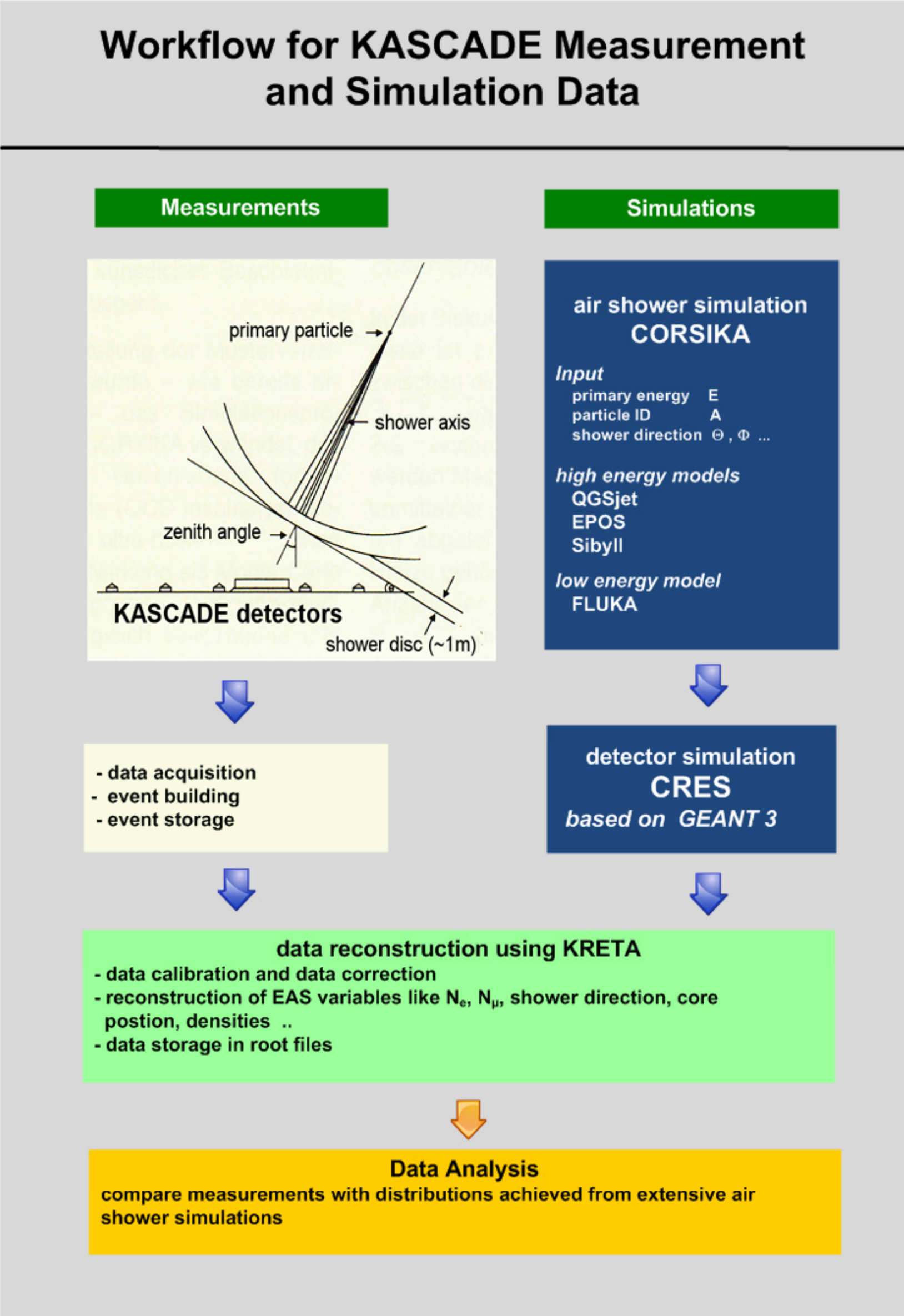}
}
\caption{KASCADE data analysis workflow for measurements and simulations.}
\label{Workflow}
\end{figure}

At KASCADE, the entire simulation chain consists of three parts:
(i) air shower simulation  performed by CORSIKA;
(ii) detector simulation performed by CRES (Cosmic Ray Event Simulation);
(iii) data reconstruction performed by KRETA.
Fig.~\ref{Workflow} illustrates the parallel workflow of measurements and simulations
as applied in KASCADE (and Grande). \\

CORSIKA is a detailed Monte Carlo program to study the evolution and
properties of extensive air showers in the atmosphere.
Protons, light nuclei up to iron, photons, and many other particles can
be treated as primaries. The particles are tracked through the atmosphere until
they undergo reactions with the air nuclei or - in the case of non-stable secondaries –
decay. A variety of high- and low-energy hadronic interaction models is implemented.
In KASCADE we were using six high energy models from three different model families
(for a comparison of the models see~\cite{tanguy} and references therein) -
QGSjet-II-02 and QGSjet-II-04;
EPOS 1.99 and EPOS-LHC;
SIBYLL 2.1 and SIBYLL 2.3; -
and one low energy model in different versions, named FLUKA.

The detector simulation is performed with CRES, a
code package for the simulation of the signals and energy deposits in all detector
components of KASCADE-Grande as response to an extensive air shower as simulated with
CORSIKA. CRES has been developed, based on the GEANT3~\cite{geant} package accepting
simulated air shower data from (unthinned) CORSIKA as input delivering simulated detector signals.
The data structure of the CRES output is the same as from the KASCADE measurements,
which means that both are analysed using the same reconstruction program KRETA.
 Unlike for measured data where we have calibration parameters like air temperature and event
specific information like the event time, we have here some additional information on the shower
properties like true primary energy and particle ID derived directly from CORSIKA or from CRES.
It was one of our main goals to publish the simulation data in the same format as the measured data
published with the release NABOO, to make it as easy as possible for the users.

From about 200 observables obtained in the analysis of the simulated data we choose 34 to be
published in KCDC.
Some of these parameters are representing the true shower information, which are described as: \\
\textbf{True primary energy:}
The energy of the particle inducing the air shower is an input for the CORSIKA air
shower simulation code.
In our case we simulated showers with a primary energy between $10^{14}\,$eV and
$3.16 \cdot {10}^{17}\,$eV following a power law spectrum with an index of $-2$.\\
\textbf{True primary particle ID:} The ID of the particle inducing the air shower
is an input for the CORSIKA air shower simulation code. We simulated 5 primaries
representing 5 different mass groups. These primaries and their respective IDs are:
\begin{table}[h]
\label{tab:Primaries}
\begin{tabular}{llrl}
\toprule
proton  & ID = & 14   & representing the lightest mass group   \\
helium  & ID = & 402  & representing a light mass group        \\
carbon  & ID = & 1206 & representing the CNO-group             \\
silicon & ID = & 2814 & representing a medium heavy mass group \\
iron    & ID = & 5626 & representing a heavy mass group        \\
\bottomrule
\end{tabular}
\vspace{-4mm}
\end{table} \\
\textbf{True shower direction:}
The zenith angle and the azimuth angle of the incident particles are input
parameters for the CORSIKA air shower simulation code.
The zenith angle spectrum reaches from $0^\circ$ to $42^\circ$ in simulation.
The zenith angle is selected at random in this interval to match equal particle fluxes
from all solid angle elements of the sky and a registration by a horizontal
flat detector arrangement. The azimuth angle is always simulated between
$0^\circ$ and $360^\circ$, where $0^\circ$ corresponds to an shower axis pointing to
the North and $90^\circ$ to the East. \\
\textbf{True numbers of electrons:} The true number of electrons is derived
from the CORSIKA output as the number of electrons tracked down to the observation
level of KASCADE at $110\,$m asl. Only electrons above $3\,$MeV low energy
cut-off are taken into account. \\
\textbf{True numbers of muons:}
The true number of muons is derived from the CORSIKA output as the number of
muons tracked down to the observation level.
Only muons above $100\,$MeV low energy cut-off are taken
into account. \\
\textbf{True numbers of photons:}
The true number of photons is derived from the CORSIKA output as the number of
photons (and $\pi^{0}$!) above $3\,$MeV tracked down to the observation level. \\
\textbf{True numbers of hadrons:}
The true number of hadrons is derived from the CORSIKA output as the number of hadrons
tracked down to the observation level and above $100\,$MeV. \\
\textbf{True shower core position:}
The true shower core position is derived from the detector simulation (CRES)
output defined as the position within the detector area where the shower centre is
located. In CRES this centre can be chosen when initializing the detector simulation.
The core positions are uniformly distributed
over an area slightly larger than the detector array, without any fiducial area cuts applied.

\subsection{Sample of cosmic-ray spectra}

With the latest release, spectra data sets from a number of experiments in the cosmic
ray field are available for download.
Currently 88 data sets are provided from 21 different experiments published between
1984 and 2017 in the energy range $E_0=10^{14}{-}{10}^{20}\,$eV. In case of KASCADE
and KASCADE-Grande we published
as well the data sets from the different mass groups
derived from the unfolding procedure for different high-energy interaction models like
QGSJet, EPOS and SIBYLL.
All data sets
are basically stored with a spectral index $\gamma=0$, but the index for the download
data can be chosen as well as the format settings. The errors given comprise only
statistical errors as published by the authors.

\begin{figure*}[ht]
\centering
\resizebox{0.79\textwidth}{!}{%
\includegraphics{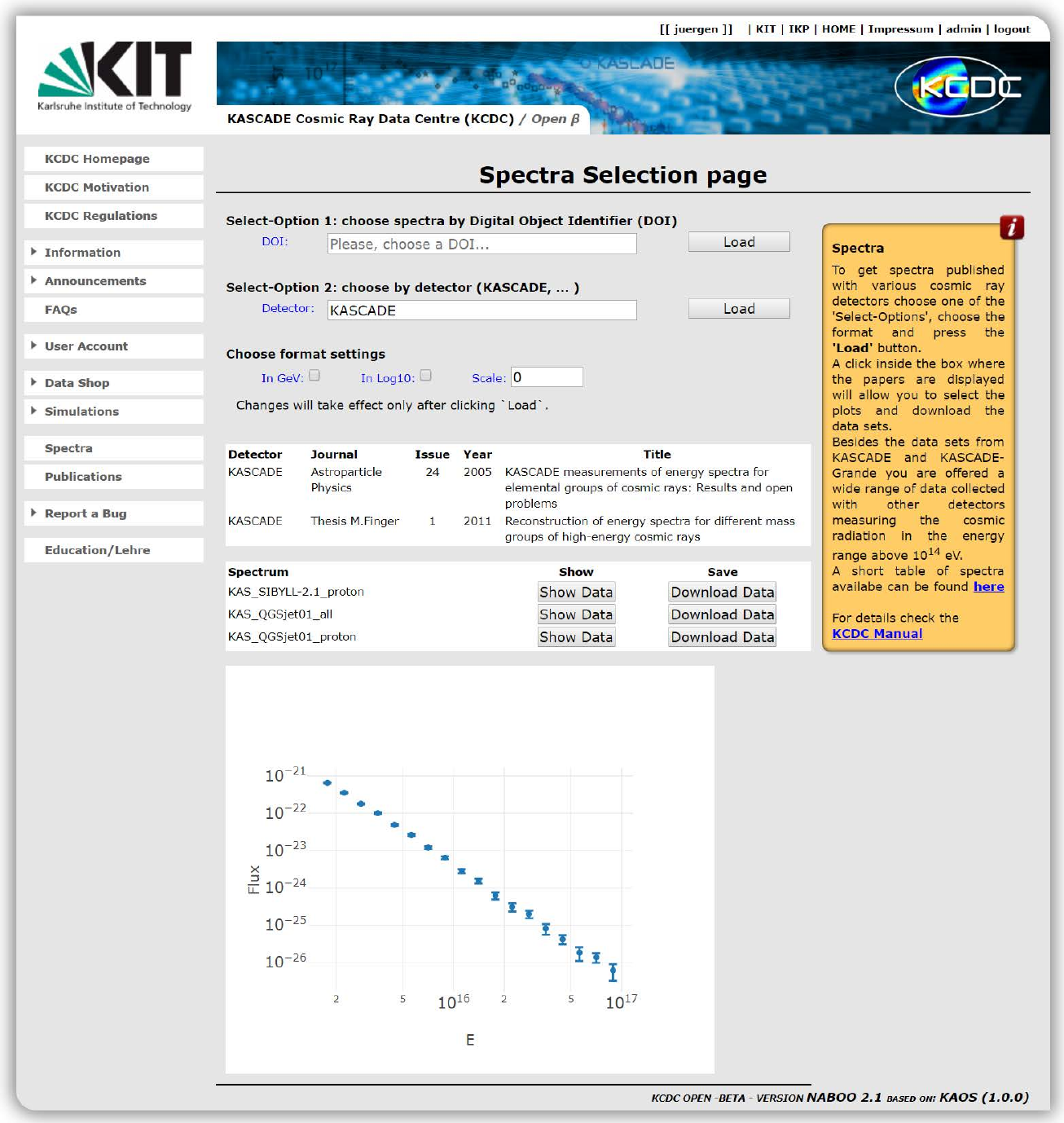}
}
\caption{Example of a spectrum ready for download.}
\label{Spectra}
\end{figure*}

Fig.~\ref{Spectra} shows an example of the KCDC spectra selection and download page.
\begin{figure*}[t]
\centering
\resizebox{0.79\textwidth}{!}{%
\includegraphics{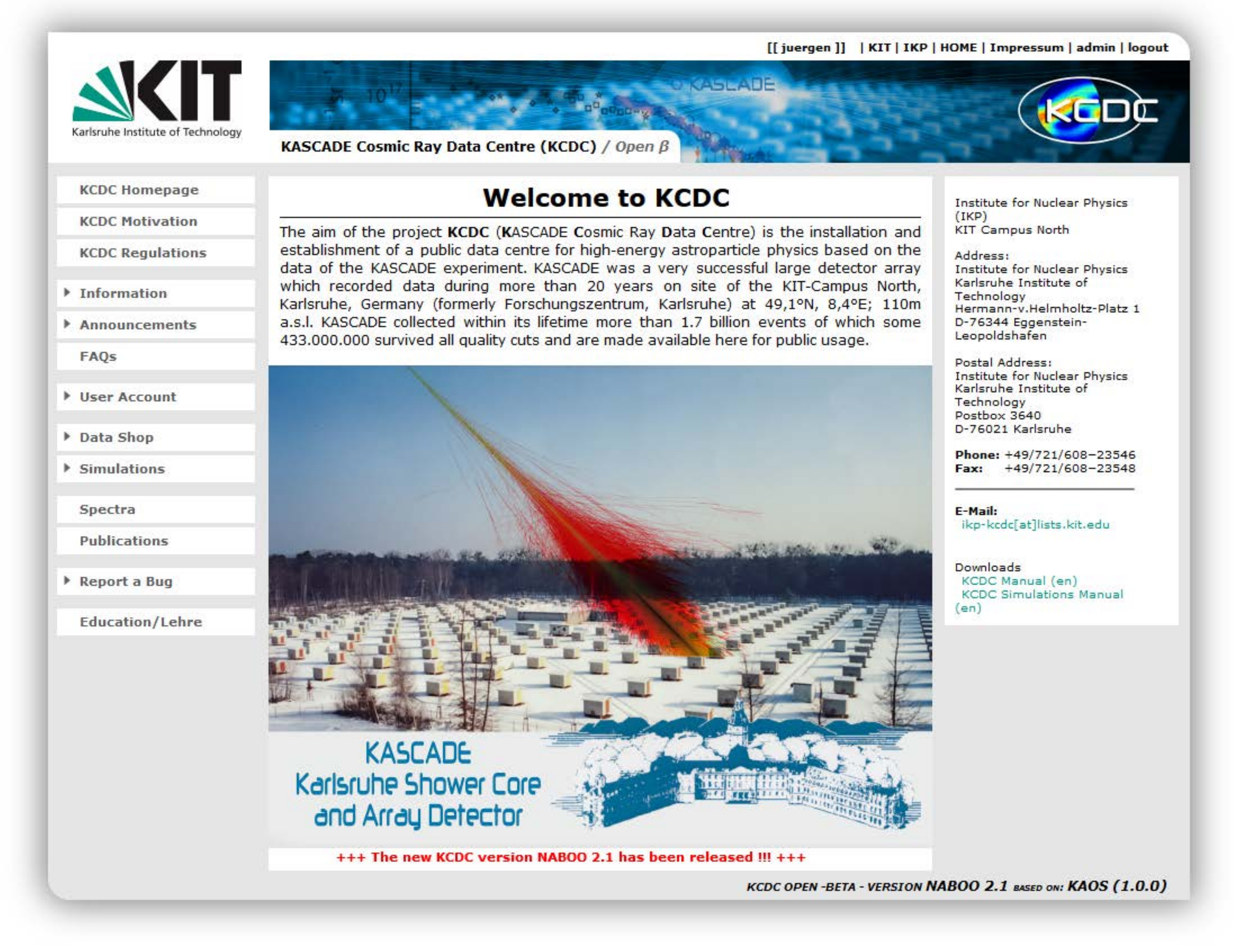}
}
\caption{Entry page of KCDC.}
\label{Homepage}
\end{figure*}

\section{The Web Portal}

The web portal (entry page see fig.~\ref{Homepage}) as interface between the data archive,
the software of the data centre and the user is the central part of KCDC.
It provides the door to the open data publication, where the baseline concept follows the
`Berlin Declaration on Open Data and Open Access'~\cite{Berlin} which explicitly requests the
use of web technologies and free, unlimited access for everyone.

We declared both, the scientific and the non-scientific audience as focus of possible users.
This requires extensive documentation of experiment, data, and software on a level
understandable and handy for all.

\begin{figure}[h]
\centering
\resizebox{0.47\textwidth}{!}{%
\includegraphics{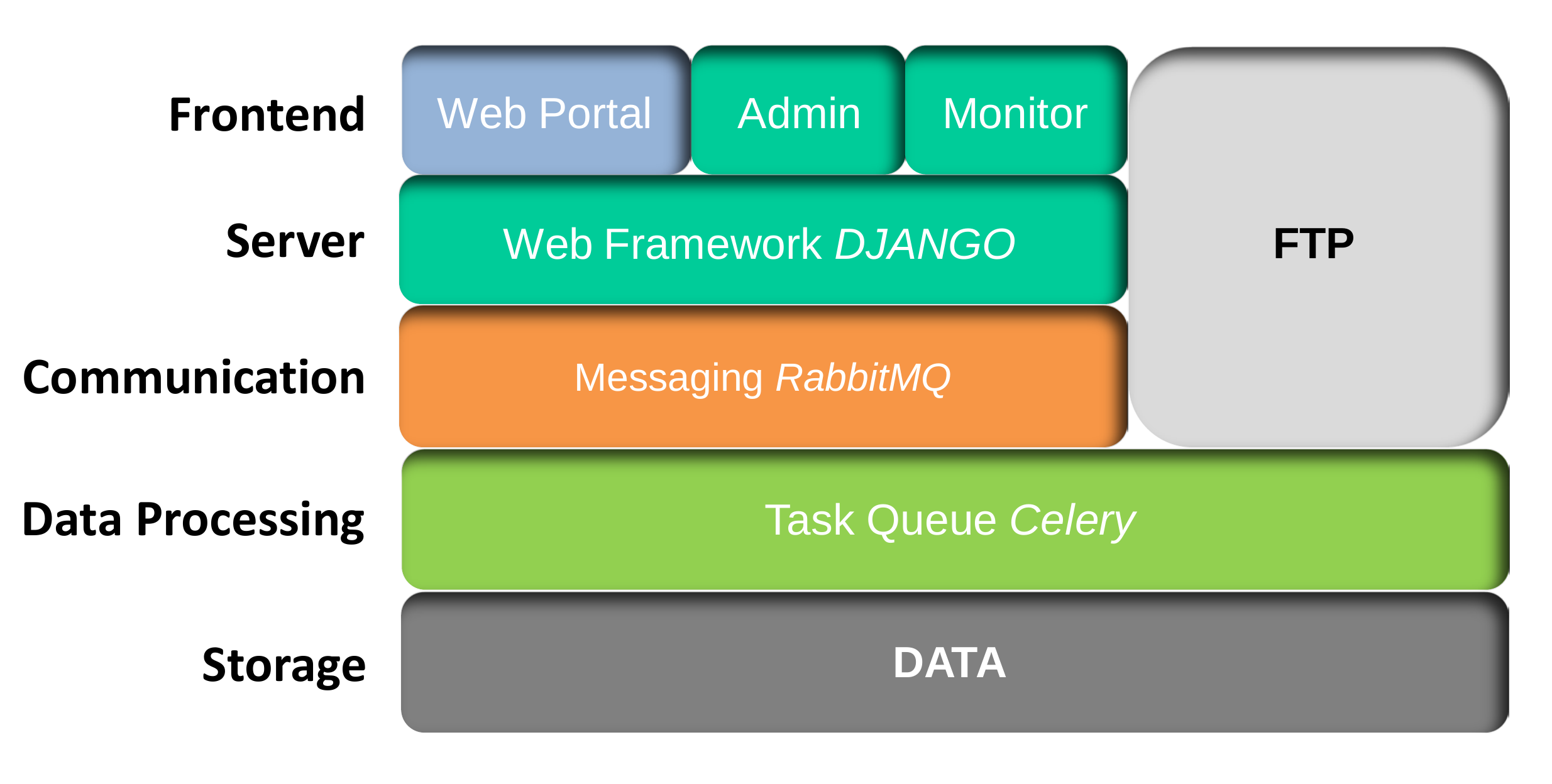}
}
\caption{IT architecture of the KCDC web portal.}
\label{KCDC-IT}
\end{figure}

\begin{figure}[h]
\centering
\resizebox{0.49\textwidth}{!}{%
\includegraphics{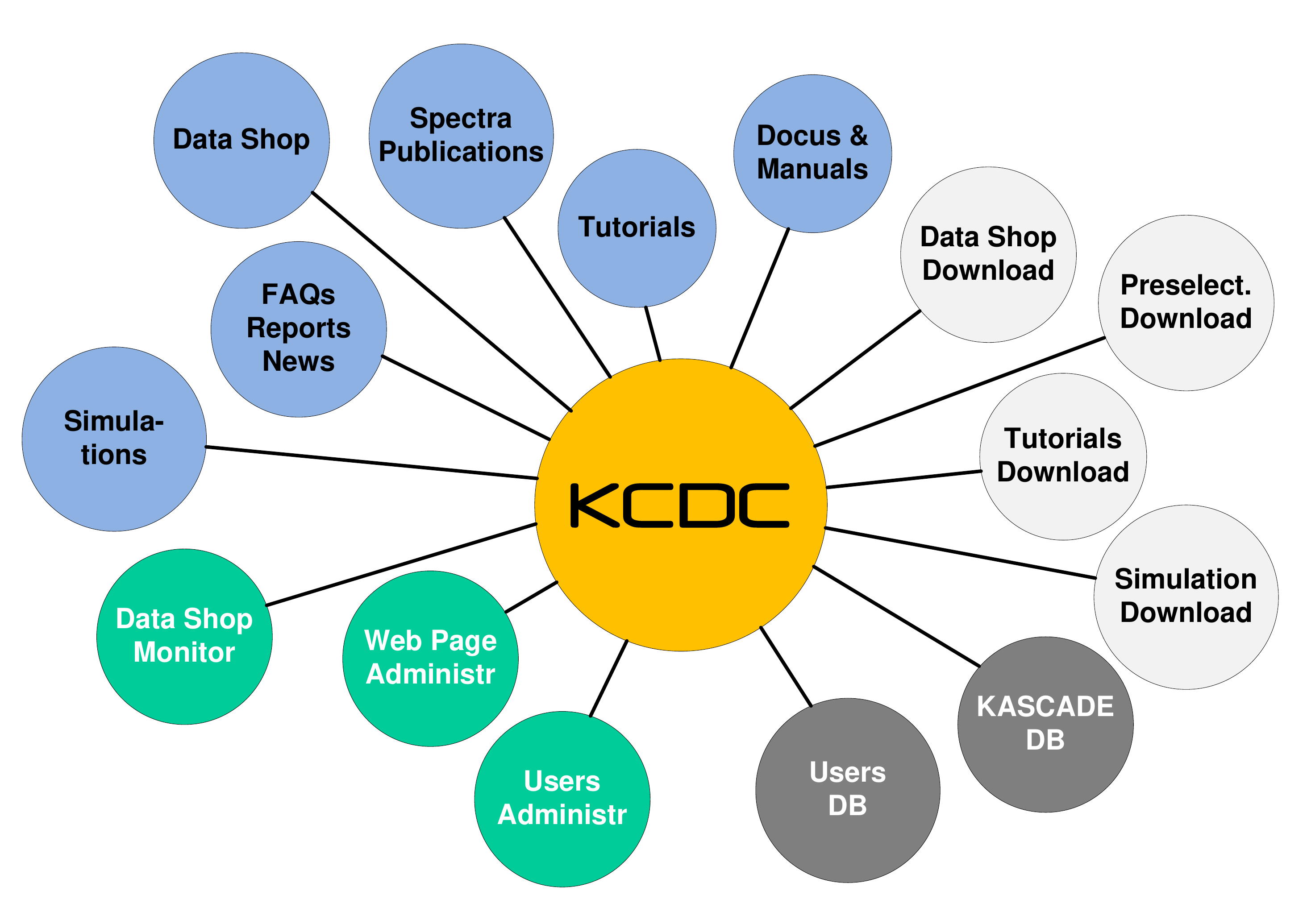}
}
\caption{Features of the KCDC web portal.}
\label{KCDC-Feat}
\end{figure}

The portal uses modern technologies, including standard internet access and
interactive data selections.
The selected data are provided for download via a corresponding FTP server.
Figs.~\ref{KCDC-IT} and~\ref{KCDC-Feat} schematically show the basic concept
of the KCDC web portal.

To serve as a general software solution for open access to (astroparticle) data,
KCDC is build as a modular, flexible framework with a good scalability
(e.g.~for installing at large computing centres).
It is foreseen that the software behind the data centre including the web portal
is also made freely available. We ensured by using modern
open source software components that not only the KASCADE-Grande data can be published,
but also other open data projects can be served with the underlying KAOS Software.
KAOS will be released as Open Source for free use.
\begin{figure*}[t]
\centering
\resizebox{0.79\textwidth}{!}{%
\includegraphics{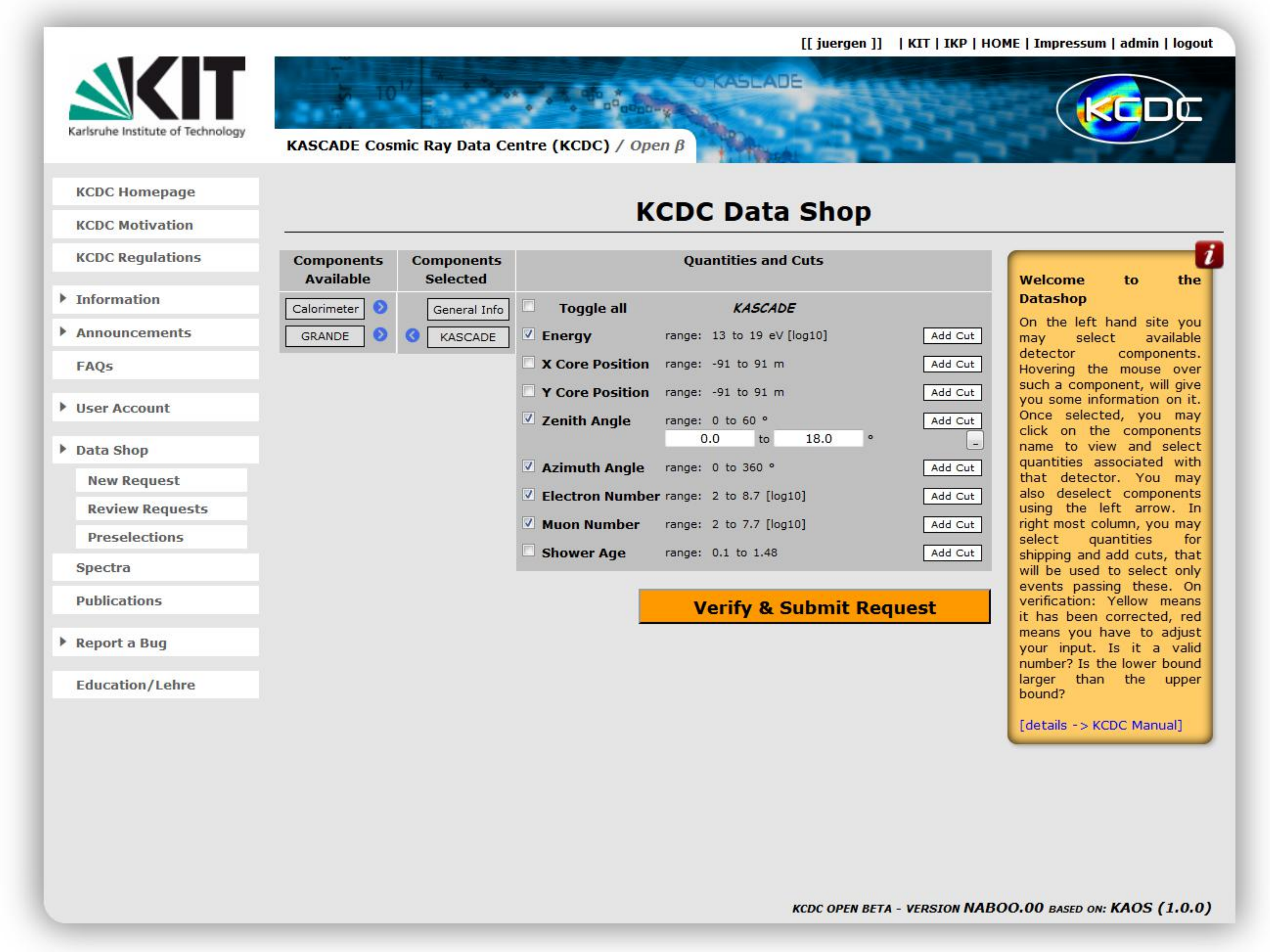}
}
\caption{KCDC data shop with detector components, quantities to be selected and
cuts to be applied on.}
\label{DataShop}
\end{figure*}

\subsection{Data availability}

Since the first release of KCDC in November 2013, where we published about
158 million events with 15 parameters each, we have widely extended our data shop.
With our latest release NABOO from February 2017  we provide more than 433 million
events from the whole measuring time of KASCADE-Grande.
Furthermore we extended the number of quantities per event from more detector
components (KASCADE, Grande and Hadron Calorimeter) plus data arrays providing
information on energy deposits and arrival times for every single detector station.

Figure~\ref{DataShop} shows a screen shot of the data shop with the list of
parameters available per event.
If registered via the `user page', the user is able to enter the data shop.
A registration is necessary in order to ensure that the
`End User License Agreement' is read, i.e.~the legal aspects of public data
are understood.
At the data shop the user can select specific event samples. For each parameter
a description is available in a corresponding info box appearing by a
mouse-over function.
A more complete information is given in the `KCDC manual', available already at the KCDC
home page.
After defining cuts the selection can be submitted.
The user gets an email notification when the
selection has been processed and is ready for download via an FTP server.
The data will be at the user's disposal in ROOT, HDF5 or ASCII-format including a
detailed header with descriptions of the selection and the data format.
Simulations can be downloaded in ROOT format only, in event packages of  different mass groups for
six high-energy interaction models.
If data arrays are selected no ASCII data format, but ROOT and/or HDF5 will be
available for download because such huge amount of data cannot be handled by ASCII
files.
Also several preselections of KASCADE data are available directly at the data shop.

There is no restriction to the user on the kind of analysis with the provided
data nor the publication of the results.
However, the KCDC team would acknowledge notification on a use exceeding
private education, as well as bug
reports or suggestions for improvements.
This can be done directly via the web portal and/or by
\email{ikp-kcdc@lists.kit.edu}.

\subsection{Legal aspects of KCDC}

The data can be used for any analysis, presumed that the user accept the
`limited use licence'.
Opposite to software open source publications, there is no standard procedure
yet available for open data publication.
In cooperation with KIT and its law department we developed an own license based on the
EULA (End User License Agreement) model~\cite{eula}, adapted from that one often used
for software (following text is taken from the KCDC-EULA):

{\it Subject to your agreement and continuing compliance with the KCDC Terms,
KIT hereby grants to you a limited, personal, nonexclusive, non-transferable,
non-assignable and fully revocable license to --
(a) use the web portal and
(b) download and use the scientific data of the KCDC in compliance with good scientific practice --
provided through the web portal or related online services for your non-commercial scientific purposes only.
Commercial purposes are defined as projects for your own or third parties for which you are paid
or granted values in lieu of cash for the use of the data.}

We had to consider a twofold issue as the license is needed for the web
portal and the data. The KCDC approach is based on the EULA model,
because it is   flexible and adaptable to our needs, it includes the idea of
requiring a good scientific practice, has to be signed
during registration and is shipped with each data package.
In our custom-made adaption of the KCDC-EULA we followed some key points from
industry, like
\begin{enumerate}      
\item no warranty for damage by owner of web portal or data;
\item no guarantee for availability or uptime of the server;
\item   in case of disputes with local laws the EULA intention is conserved;
\item changes are possible at any time;
\item the termination of EULA is at our digression, only,
\end{enumerate}
as well as obvious requirements from the open data idea, like
\begin{enumerate}      
\item free access to the data and the web portal;
\item good scientific practice for the work with the data;
\item commercial usage of the data is not prohibited\footnote{Please note, that the apparent contradiction to the statement above concerning the non-commercial use of the data, is solved by following statement in the EULA: `As an alternative to this EULA, KIT offers a license to use the DATA commercially as well, on the basis of a commercial license agreement. If YOU are interested in such a license please contact KIT, Institute for Nuclear Physics (IKP), the KCDC Group and the contact person provided on the WEBPORTAL'.};
\item the citation of the collaboration, the KIT, and the KCDC web portal is mandatory;
\item free redistribution of data `as is'.
\end{enumerate}

\subsection{Tutorials}

\begin{figure*}[ht]
\centering
\resizebox{0.79\textwidth}{!}{%
\includegraphics{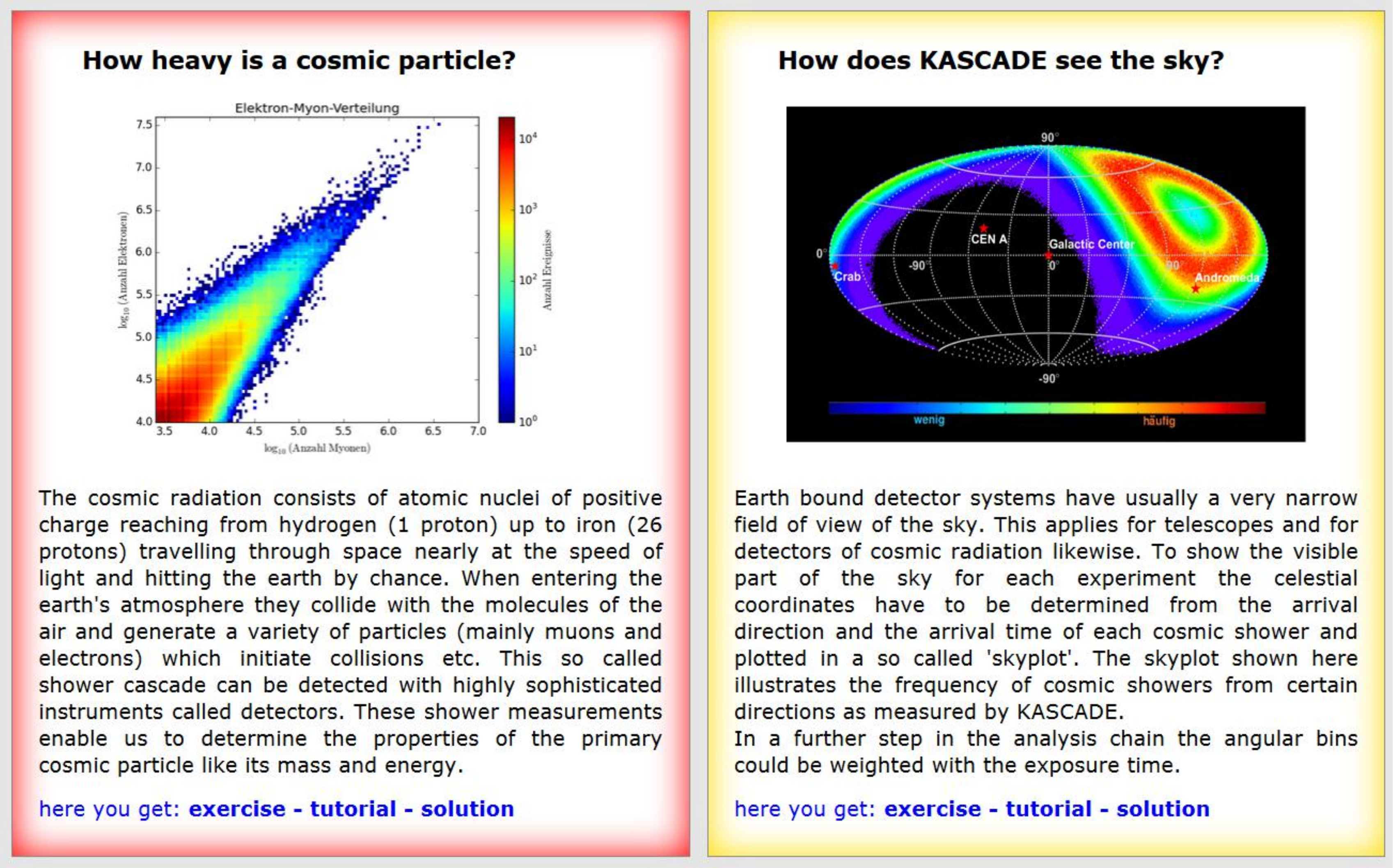}
}
\caption{Impression of the tutorial section of KCDC.}
\label{Tutorial}
\end{figure*}

The goal of having detailed tutorials, i.e.~an `education portal', is to
provide the data also to a general public in the sense of a visible outreach of
astroparticle physics. The main addressees are pupils and students interested
in this research field. This implies that a tutorial
has to provide a basic knowledge on KASCADE, astroparticle physics and
related topics, the software necessary to analyze the data and the KCDC data,
preferably as preprocessed data selection.
Furthermore, a step-by-step explanation of a simple analysis and a detailed
discussion of the results are necessary together with a code fragment as example.
Up to now only a few tutorials are available in German and in English. More tutorials
also in other languages like French can be added without any problem and will
be done in the near future, developed together with teachers and pupils. We also
invite the KCDC users to contribute to the education portal.
Figure~\ref{Tutorial} shows a screen shot of the KCDC education page with
available exercises.

\subsection{Analyses with KCDC data sets}
\label{kascade-analysis-for-the-datasets-and-physics-prospects}

A general idea of KCDC is to allow a repetition of all the published analyses of the
KASCADE-Grande experiment (also on basis of newly developed hadronic
interaction models). In particular, due to the constant advancement of the
hadronic interaction models it is worthwhile to cross-check the validity of these models
also in future with the high-quality data of KASCADE-Grande.
This, of course, can be done only by physicist on an expert level,
but via KCDC, not only by internal collaboration members.

We have checked this personal expectation by performing some of the analyses
based on the information of KCDC, only.
With the release of the data sample, many quantities are now
publicly available. While two quantities (run and event IDs) serve
solely for comparison purposes, the remaining 25 observables can be used
for a wide range of physics analyses at different levels.

At a basic level the data can be used by teachers in
physics courses to kindle an interest in astroparticle physics and
cosmic rays in particular. For example, the simple energy estimate for each event,
which is included in the published data-sample, can be used to show the
energy dependence of the number of muons and electrons that are produced
in an air shower and reach the observation level. The direction and time
of the arrival of the cosmic particles is all that is needed to produce
a skymap, which could serve as an application for coordinate
transformations and different projections.

Another example is to study the influence of the atmospheric conditions
such as the temperature and
air pressure on the shower development. Under the
assumption that the composition and intensity of cosmic rays at a
certain primary energy are independent of the arrival direction, the
zenith angle can be used to vary the amount of atmosphere that the
shower needs to pass. One application could be to study the attenuation
of muons and/or electrons in the atmosphere. Since the KASCADE array is
located at around $110\,$m above sea level the atmospheric depth is already
high for vertical showers. Colleagues with access to comparable data of
similar experiments at higher altitudes could use the KASCADE data to
extend the accessible range of traversed atmosphere.

Many more examples can be given, but our goal to reach all target groups -
cosmic ray physicists; public users (physicists); teachers, pupils,
early researchers, e.g. - was reached and proven by our monitoring and the
feedback by the users.

\section{Future Steps}

In light of recent international efforts of strengthening, even
enforcing, open access in science especially for science data and on the
other hand the lack of adequate methods readily available both in the
cosmic ray and astroparticle physics community, the need of a reliable
solution for open data publications is growing. We believe that the
presented solution based on a web portal will provide a
flexible and adaptable tool for other scientists within and outside the
physics community to do open data publications of their collected data
with as little effort as possible.

We have several tasks and ideas on our to-do list for the further development
of KCDC. For the data shop next steps we intend to publish data from the
radio antenna array LOPES which was run in coincidence with KASCADE for more
than 8 years.
Furthermore we intend to offer the data from the combined analysis of the
KASCADE and Grande detectors which equates to a new detector in the KCDC data shop.

In addition, one of the main goals is to extend the educational portal
with more examples to be more attractive for users on high-school level.

On the technical side we will publish the software
(KAOS) to encourage other astroparticle
physics experiments to provide their data.

Last but not least we want to enhance the visibility of KCDC by
including it to general data catalogs and long-term data archive networks, similar to,
e.g. Re3data~\cite{re3data}, where KCDC is already present. \\

In parallel and as a long-term goal we want to use KCDC as basis of
a global `Analysis \& Data Centre in Astroparticle Physics' which allows not
only for data access, but also provides resources for the analysis
of the data. Main motivation is here that astroparticle physics requests
more and more for multi-messenger analyses, which will need an experiment-overarching
platform. The high demand in the German and
international community as well as that our observatories are globally
distributed request an open science system which can be based on KCDC and
the Tier centre capabilities of particle physics as well as on
experiences of the commonplace virtual observatories in astronomy.
This will need the development of integrated solutions of distributed data
storage algorithms and techniques to allow the community to perform multi-messenger
analyses, e.g.~with deep learning methods.

Towards such a data and analysis centre we are working on the KCDC
extension by scientific data from other experiments, allowing on-the-fly
multi-messenger data analyses.
The extension of KCDC in respect of the inclusion of Big Data Science Software
will allow not only access to the data, but also the possibility of developing
specific analysis methods and corresponding simulations in one environment.
This needs a move of KCDC to most modern computing, storage and data access
concepts.

\section*{Notice}
Please note that since this article has been written, KCDC was further developed and improved.
The description here is based on the {\it Open Beta Version: NABOO.2}. Please visit the KCDC
page (\url{https://kcdc.ikp.kit.edu}) and click the button `Announcements - Developer News'
for more information.

\section*{Acknowledgements}
The KCDC team within the KASCADE-Grande collaboration acknowledges the continuous support of the
project by the `Helmholtz Alliance for Astroparticle Physics HAP' funded by the Initiative and
Networking Fund of the Helmholtz Association. In addition we acknowledge the fruitful cooperation with
the Steinbuch Computing Centre at KIT as well as the cooperation with the local Thomas-Mann-Gymnasium,
Blankenloch and its interested pupils and with students from the University Strasbourg.



\begin{thebibliography}{}
%
\bibitem{Haungs:2015zna}
 A.~Haungs et al. (KASCADE-Grande Collaboration), J.\ Phys.\ Conf.\ Ser.\  {\bf 632},  012011 (2015)
\bibitem{kascade}
T.~Antoni et al.  (KASCADE Collaboration),  Nucl.\ Instr.\ Meth.\ A {\bf 513}, 490 (2003)
\bibitem{grande}
W.D.~Apel et al. (KASCADE-Grande Collaboration), Nucl.\ Instr.\ Meth.\ A {\bf 620}, 202 (2010)
\bibitem{lopes}
H.~Falcke et al. (LOPES Collaboration), Nature {\bf 435}, 313 (2005)
\bibitem{crome}
R.~Smida et al. (CROME Collaboration),  Eur. Phys. J. Web Conf.  {\bf 53}, 08010 (2013)
\bibitem{calorimeter}
J.~Engler  et al., Nucl.\ Instr.\ Meth.\ A {\bf 427},  528 (1999)
\bibitem{myontrackdet}
P.~Doll et al., Nucl.\ Instr.\ Meth.\ A {\bf 488},  517 (2002)
\bibitem{Apel2012183}
W.D.~Apel et al. (KASCADE-Grande Collaboration), Astropart. Phys. \textbf{36}, 183 (2012)
\bibitem{kas-unf}
T.~Antoni et al. (KASCADE Collaboration), Astropart. Phys. \textbf{24}, 1 (2005)
\bibitem{haungs-review}
A.~Haungs, H.~Rebel, M.~Roth, Rept.\ Prog.\ Phys. {\bf 66}, 1145, (2003)
\bibitem{KGhaungs-ICRC15}
A.~Haungs et al. (KASCADE-Grande Collaboration), ICRC 2015, The Netherlands, PoS(ICRC2015)278
\bibitem{prl107}
W.-D.~Apel et~al. (KASCADE-Grande Collaboration), PRL \textbf{107}, 171104 (2011)
\bibitem{Apel2013}
W.-D.~Apel et~al. (KASCADE-Grande Collaboration), PRD \textbf{87}, 081101(R) (2013)
\bibitem{corsika}
D.~Heck et al., report Forschungszentrum Karlsruhe, FZKA \textbf{6019}, (1998)
\bibitem{hadrint}
T.~Antoni et al. (KASCADE Collaboration),  J. Phys. G  \textbf{36}, 035201 (2009)
\bibitem{mtdint}
W.-D.~Apel et al.  (KASCADE-Grande Collaboration), Astrop. Phys. \textbf{65}, 55 (2015)
\bibitem{kcdc}
S.~Schoo et al.  (KASCADE-Grande Collaboration), ICRC 2015, The Netherlands, PoS(ICRC2015)262
\bibitem{kcdc-icrc17}
D.~Kang et al. (KASCADE-Grande Collaboration), ICRC 2017, Busan, PoS(ICRC2017)452
\bibitem{Berlin}
see \url{https://openaccess.mpg.de/Berlin-Declaration}, January (2015)
\bibitem{refFAIR}
M.D.~Wilkinson et al., SCIENTIFIC DATA  \textbf{3}, 160018 (2016)
\bibitem{qgsjet}
S.~Ostapchenko,  PRD, \textbf{83}, 014018 (2011)
\bibitem{fluka}
A.~Fass\`o, et al., Report CERN-2005-10, INFN/TC-05/11, SLAC-R-773 (2005)
\bibitem{tanguy}
T.~Pierog, JPS Conf. Proc. \textbf{19}, (2018) 011018
\bibitem{geant}
R.~Brun et al., GEANT 3 User Guide, CERN/DD/EE/84-1 (1987)
\bibitem{eula}
see \url{http://en.wikipedia.org/wiki/End-user_license_agreement}, January 2015
\bibitem{re3data}
see \url{http://www.re3data.org/}, January 2015

\end{thebibliography}
\end{document}